\documentclass[aps,prl,reprint,superscriptaddress,amsmath,amssymb]{revtex4-2}

\usepackage{graphicx}
\usepackage{txfonts}
\usepackage[hypertexnames=false]{hyperref}
\usepackage{braket}
\usepackage{bm}
\usepackage{enumitem}

\begin{document}

\title{Mean-Field Convective Phase Separation under Thermal Gradients}

\author{Meander Van den Brande}
\affiliation{Department of Mathematics, King’s College London, Strand, London WC2R 2LS, United Kingdom}

\author{François Huveneers}
\affiliation{Department of Mathematics, King’s College London, Strand, London WC2R 2LS, United Kingdom}

\author{Kyosuke Adachi}
\affiliation{RIKEN Center for Interdisciplinary Theoretical and Mathematical Sciences, 2-1 Hirosawa, Wako 351-0198, Japan}
\affiliation{Nonequilibrium Physics of Living Matter Laboratory, RIKEN Pioneering Research Institute, 2-1 Hirosawa, Wako 351-0198, Japan}

\date{\today}

\begin{abstract}

Nonequilibrium conditions fundamentally change how systems undergo phase separation.
In lattice gases with attractive interactions under temperature gradients, particles have been shown to form periodic patterns and steady convective currents; however, a clear theoretical explanation of this convective phase separation is still missing.
Here, we present a dynamical mean-field model that describes the mechanism behind this convective phase separation.
Using linear stability analysis, we show that the transition from a uniform state to a periodic pattern is driven by the emergence of a dominant unstable mode.
Numerical simulations confirm the predicted phase diagram and demonstrate that these convective currents are a robust feature of the steady state, appearing regardless of the initial conditions.
These results provide a direct approach for understanding how temperature gradients drive the formation of steady-state convective patterns.

\end{abstract}

\maketitle

\paragraph{Introduction.}

Pattern formation occurs widely in both equilibrium and nonequilibrium systems~\cite{Cross1993}.
In equilibrium, phase ordering generates spatial structure via a coarsening process that has been extensively described theoretically~\cite{Cahn1958, Kawasaki1966, Koch1983, Bray1994, Onuki1997} and observed experimentally in polymer solutions and blends~\cite{Tanaka2000, Rubinstein2003, Cabral2018}.
Typically, attractive interactions drive the system toward a universal growth law~\cite{Bray1994}, ultimately resulting in macroscopically large domains~\cite{Hohenberg1977}.

Nonequilibrium conditions can fundamentally alter the mechanisms and morphology of phase separation by breaking detailed balance.
A prominent example is found in active matter, where self-propulsion leads to collective motion and motility-induced phase separation~\cite{Vicsek2012, Chate2020, Fily2012, Cates2015}, potentially exhibiting persistent microphase separation~\cite{Tjhung2018, Shi2020, Cates2025}.
Similar departures from equilibrium ordering occur in systems driven by chemotaxis~\cite{Keller1970, Weyer2025}, catalysis~\cite{Cotton2022}, shear flow~\cite{Onuki1997, Corberi1998, bbhy-gmsp}, or external driving forces~\cite{Katz1984, Schmittmann1995, Marro1999}.

A particularly intriguing phenomenon has recently been observed in attractive lattice gases under temperature gradients, where the system undergoes periodically patterned rather than macroscopic phase separation~\cite{Pleimling2010, Li2012, MS1} (see also \cite{Ball1990, Furukawa1992, Alt1992, Araki2004, Jaiswal2013}).
These states are sustained by convective currents reminiscent of the convection cells in classical fluid dynamics, represented by Rayleigh--B\'enard convection~\cite{Chandrasekhar1961, Ahlers2009}.
In contrast to such convection, which requires gravity, the circulating currents in the lattice gases emerge from diffusive transport, providing a minimal setting for convective pattern formation.
Despite this numerical evidence, a theoretical framework for this behavior is lacking.
Specifically, it is unclear if a deterministic model can capture these features and provide an analytical perspective, similar to the framework for Turing instabilities~\cite{Turing1952, Murray2002}.

\begin{figure*}[t]
    \centering
    \includegraphics[scale=1]{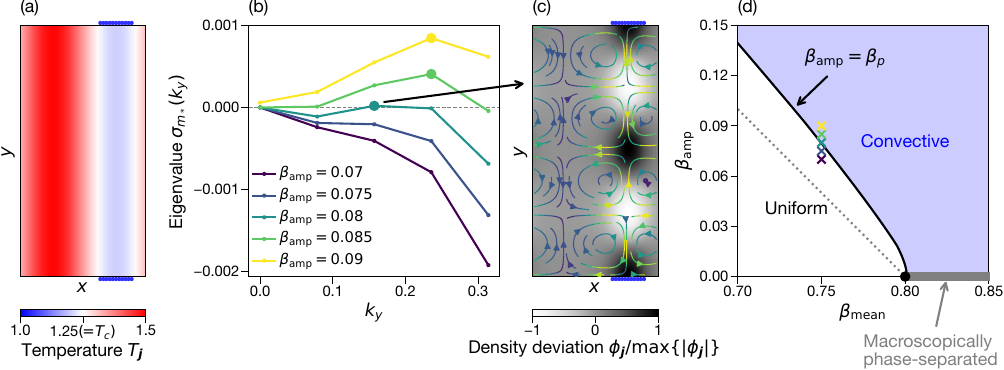}
    \caption{Linear instability: periodic density modulation and convective currents.
    (a) Temperature profile in Eq.~\eqref{eq_temperature_profile_example} for $\beta_\mathrm{mean}=0.75$ and $\beta_\mathrm{amp} = 0.08$.
    Blue dots indicate $T_{\bm{j}} < T_c$.
    (b) Dispersion relation: largest eigenvalue $\sigma_{m_*} (k_y)$ of the linearized dynamics in Eq.~\eqref{eq_linearized_dynamics} near the stability limit for $\beta_\mathrm{mean} = 0.75$ and various $\beta_\mathrm{amp}$.
    (c) Most unstable eigenvector for $\beta_\mathrm{mean}=0.75$ and $\beta_\mathrm{amp} = 0.08$:
    density modulation (grayscale) and current pattern (colored lines, brightness indicates magnitude).
    (d) Phase diagram based on linear stability.
    White (blue) regions indicate stability (instability) of the uniform density state toward convective patterns.
    Colored crosses denote parameters for panel (b).
    The thick gray line and black dot mark the unstable region and critical point for macroscopic phase separation at homogeneous temperature.
    The gray dotted line corresponds to $\beta_\mathrm{mean} + \beta_\mathrm{amp} = 1/T_c$ (i.e., $\min_{\bm{j}} T_{\bm{j}} = T_c$).
    For all panels, $(L_x, L_y) = (40, 80)$ and $\bar{\rho} = 1 / 2$.}
    \label{fig_linear}
\end{figure*}

In this work, we introduce a deterministic mean-field counterpart to the stochastic model of Ref.~\cite{MS1} to describe phase separation under spatially inhomogeneous, slowly varying temperatures.
Using linear stability analysis, we identify the most unstable mode that determines the phase boundary and the periodic convective patterns (see Fig.~\ref{fig_linear}).
We numerically validate the predicted phase diagram and demonstrate that while the final pattern selection is sensitive to initial conditions, the convective currents emerge as a robust signature of the steady state.
Finally, we show that this deterministic approach captures the essential physics of the stochastic model in Ref.~\cite{MS1}.

\paragraph{Model.}
We study a density field $\rho_{\bm{j}}(t) \in [0,1]$ on a rectangular lattice of size $L_x \times L_y$ with periodic boundary conditions and spatially inhomogeneous temperature $T_{\bm{j}}$.
We assume throughout that $T_{\bm{j}}$ varies on a macroscopic scale, $T_{\bm j} = T(j_x/L_x,j_y/L_y)$;
see Fig.~\ref{fig_linear}(a) and Eq.~\eqref{eq_temperature_profile_example} for the example considered throughout.
Here, $\bm{j}=j_x \bm{e}_x + j_y \bm{e}_y$ denotes the lattice coordinate, with $j_a = 0,1,\cdots,L_a-1$ and $\bm{e}_a$ the unit lattice vectors along the $a$ direction for $a=x,y$.
For example, $\rho_{\bm{j} - \bm{e}_x}$ below denotes the density at the left nearest neighbor of site $\bm{j}$, while $\rho_{\bm{j} + \bm{e}_x + \bm{e}_y}$ denotes that at the upper-diagonal next-nearest neighbor of $\bm{j}$, and so on.

Measuring length in units of the lattice constant and temperature in units of the interaction strength, the model reads
\begin{equation}
   \partial_t \rho_{\bm{j}} = -(J^x_{\bm{j}} - J^x_{\bm{j} - \bm{e}_x} + J^y_{\bm{j}} - J^y_{\bm{j} - \bm{e}_y}),
   \label{eq_model}
\end{equation}
where the density current $J_{\bm{j}}^a$ is given by
\begin{equation}
    J^a_{\bm{j}} = \rho_{\bm{j}} - \rho_{\bm{j} + \bm{e}_a} + (\rho_{\bm{j}} + \rho_{\bm{j} + \bm{e}_a} - 2 \rho_{\bm{j}} \rho_{\bm{j} + \bm{e}_a}) \tanh \Big( \delta^a_{\bm{j}} / (2 T_{\bm{j} + \bm{e}_a / 2}) \Big)
    \label{eq_model_current}
\end{equation}
with $\delta^a_{\bm{j}}$ the difference in the surrounding density between $\bm{j}$ and $\bm{j} + \bm{e}_a$:
\begin{equation}
\begin{aligned}
   \delta^x_{\bm{j}}
   & =
   \rho_{\bm{j} + 2 \bm{e}_x}
   + \rho_{\bm{j} + \bm{e}_x + \bm{e}_y}
   + \rho_{\bm{j} + \bm{e}_x - \bm{e}_y}
   -
   (\rho_{\bm{j} + \bm{e}_y} + \rho_{\bm{j} - \bm{e}_y} + \rho_{\bm{j} - \bm{e}_x}), \\
   \delta^y_{\bm{j}}
   & =
   \rho_{\bm{j} + 2 \bm{e}_y}
   + \rho_{\bm{j} + \bm{e}_y + \bm{e}_x}
   + \rho_{\bm{j} + \bm{e}_y - \bm{e}_x}
   -
   (\rho_{\bm{j} + \bm{e}_x} + \rho_{\bm{j} - \bm{e}_x} + \rho_{\bm{j} - \bm{e}_y}).
   \label{eq_model_delta}
\end{aligned}
\end{equation}

Equation~\eqref{eq_model} is a continuity equation, and the mean density $\bar{\rho} := (L_x L_y)^{-1} \sum_{\bm{j}} \rho_{\bm{j}}$ is conserved.
Note that $J^a_{\bm{j}}$ denotes the current on the bond from site $\bm{j}$ to site $\bm{j} + \bm{e}_a$, so that the right-hand side of Eq.~\eqref{eq_model} represents the net incoming current $\sum_{a} (J^a_{\bm{j} - \bm{e}_a} - J^a_{\bm{j}})$ at site $\bm{j}$; the apparent asymmetry between $\bm{j} - \bm{e}_a$ and $\bm{j} + \bm{e}_a$ merely reflects labeling each bond current $J^a_{\bm{j}}$ by its tail site $\bm{j}$.
In the expression for the current in Eq.~\eqref{eq_model_current}, the linear terms represent normal diffusion in the form of Fick's law, and the nonlinear one induces attractive interactions between particles, which can overcome the normal diffusion terms and lead to phase separation at low temperature.
See Sec.~\ref{app_MFA} of the Supplemental Material (SM) for the derivation of Eqs.~\eqref{eq_model} and \eqref{eq_model_current} from the stochastic model of Ref.~\cite{MS1} by neglecting microscopic density correlations.

\paragraph{Linear stability analysis.}

We analyze the linear stability of the uniform density state $\rho_{\bm{j}} (t) = \bar{\rho}$, which is always a stationary solution of Eqs.~\eqref{eq_model} and \eqref{eq_model_current}.
Linearizing Eq.~\eqref{eq_model_current} with respect to density perturbation $\phi_{\bm{j}} (t) = \rho_{\bm{j}} (t) - \bar{\rho}$, we obtain the linearized current:
\begin{equation}
     J^a_{\bm{j}} =  \phi_{\bm{j}} - \phi_{\bm{j} + \bm{e}_a} + \bar{\rho} (1 - \bar{\rho}) \delta^a_{\bm{j}} / T_{\bm{j} + \bm{e}_a / 2}.
     \label{eq_linearized_current}
\end{equation}
Introducing the Fourier transform of the inverse temperature, $\hat{\beta}_{\bm{k}} := \sum_{\bm{j}} (1 / T_{\bm{j}}) e^{-i \bm{k} \cdot \bm{j}}$, and of the density, $\hat{\phi}_{\bm{k}} = \sum_{\bm{j}} \phi_{\bm{j}} e^{-i \bm{k} \cdot \bm{j}}$,
where $\bm{k} = k_x \bm{e}_x + k_y \bm{e}_y$ and $k_a = 2 \pi l_a / L_a \in [-\pi, \pi)$ with $l_a \in \mathbb{Z}$, Eq.~\eqref{eq_model} becomes in linear approximation
\begin{equation}
     \partial_t \hat{\phi}_{\bm{k}} = \sum_{\bm{q}} R_{\bm{k} \bm{q}} \hat{\phi}_{\bm{q}}
     = (\mathbf  R\hat\phi)_{\bm k},
     \label{eq_linearized_dynamics}
\end{equation}
where
\begin{align}
    R_{\bm{k} \bm{q}} & = -2 (2 - \cos k_x - \cos k_y) \delta_{\bm{k} \bm{q}} \nonumber \\
    & \quad + \frac{4 \bar{\rho} (1 - \bar{\rho}) \hat{\beta}_{\bm{k} - \bm{q}}}{L_x L_y} \bigg[ \sin \frac{k_x}{2} \bigg( \sin \frac{3 q_x}{2} + 2 \sin \frac{q_x}{2} \cos q_y \bigg) \nonumber \\
    & \quad + \sin \frac{k_y}{2} \bigg( \sin \frac{3 q_y}{2} + 2 \sin \frac{q_y}{2} \cos q_x \bigg) \bigg].
    \label{eq_Rkq}
\end{align}
Note that both $\bm{k}$ and $\bm{q}$ label Fourier modes of the density field: $R_{\bm{k} \bm{q}}$ describes how the density mode $\bm{q}$ couples to the density mode $\bm{k}$, and the inverse temperature enters through its Fourier component at the difference wavenumber, $\hat{\beta}_{\bm{k} - \bm{q}}$.

If the temperature is uniform ($T_{\bm{j}} = T$),
the model described by Eqs.~\eqref{eq_model} and \eqref{eq_model_current} reduces to a mean-field equilibrium lattice gas model with attractive interactions as derived previously~\cite{Penrose1991}, and captures isotropic phase separation dynamics as observed in similar models with conservation laws~\cite{Oono1988, Bray1994}.
This is confirmed by linear stability analysis:
$\hat{\beta}_{\bm{k}} = L_x L_y \delta_{\bm{k} \bm{0}} / T$ and $R_{\bm{k} \bm{q}} = r_{\bm{k}} \delta_{\bm{k} \bm{q}}$,
hence, the uniform density state is linearly unstable if $\max_{\bm k}r_{\bm k} > 0$.
Expanding $r_{\bm k}$ to the second order in $\bm k$ yields
\begin{equation}\label{eq_rk_expansion}
     r_{\bm{k}} \approx [-1 + 5 \bar{\rho} (1 - \bar{\rho}) / T] \bm{k}^2 \quad (\mathrm{for} \ \bm{k} \approx \bm{0}).
\end{equation}
The linear instability appears for $T < T_s = 5 \bar{\rho} (1 - \bar{\rho})$ (spinodal temperature).
The critical density that maximizes $T_s$ is given by $\bar{\rho} = \bar{\rho}_c = 1 / 2$ and the corresponding critical temperature is $T_c = T_s |_{\bar{\rho} = \bar{\rho}_c} = 5 / 4$.
Further, the most unstable mode $\bm k_*$, defined through $r_{\bm k_*} = \max_{\bm k} r_{\bm k}$, satisfies $\bm k_* \approx \bm 0$ near $T = T_s$, which corresponds to macroscopic phase separation (see Sec.~\ref{app_EPS} of the SM).
Note that the most unstable mode here is determined from the full growth rate $r_{\bm{k}}$ [Eq.~\eqref{eq_homogeneousT_rk} in the SM], while the second-order expansion \eqref{eq_rk_expansion} is used only to locate the spinodal temperature $T_s$.

The situation changes drastically when we move to an inhomogeneous temperature profile.
The uniform density state is unstable if $\max_n \mathrm{Re}\lambda_n > 0$, where the maximum runs over all eigenvalues $\lambda_n$ of $\mathbf R$ defined in Eq.~\eqref{eq_linearized_dynamics}.
The most unstable mode $n_*$, which corresponds to the eigenvalue $\lambda_{n_*}$ with the largest real part, determines the type of instability near the stability limit, at which $\mathrm{Re} \, \lambda_{n_*} \to 0$.

From now on, let us focus on the temperature profile in Fig.~\ref{fig_linear}(a) that varies only along the $x$-axis:
\begin{equation}
    1 / T_{\bm{j}} = \beta_\mathrm{mean} - \beta_\mathrm{amp} \sin (2 \pi j_x / L_x).
    \label{eq_temperature_profile_example}
\end{equation}
Here, $\beta_\mathrm{mean}$ and $\beta_\mathrm{amp}$ represent the mean and amplitude of the inverse temperature, respectively ($\beta_\mathrm{mean} > \beta_\mathrm{amp} \geq 0$).
The temperature is highest along the vertical line $j_x = L_x / 4$ and lowest along the vertical line $j_x = 3 L_x / 4$.
The matrix elements of $\mathbf R$ in \eqref{eq_Rkq} take the form
\begin{equation}\label{eq_S_matrix}
    R_{\bm{k} \bm{q}} = S_{k_x q_x}(k_y) \, \delta_{k_y q_y}.
\end{equation}
For a given $k_y$, we write the eigenvalues of $\mathbf{S}(k_y)$ as $\sigma_m (k_y)$ with $m=1,\dots,L_x$, and the largest one as $\sigma_{m_*}(k_y)$ (which turns out to be real).
Note that the conservation of the mean density forces only the eigenvalue associated with the spatially uniform mode to vanish; the maximized growth rate $\sigma_{m_*}(k_y)$ can thus be positive even at $k_y = 0$ above the onset of instability, reflecting unstable modes modulated purely along the $x$ axis [see the curve for $\beta_\mathrm{amp} = 0.09$ in Fig.~\ref{fig_linear}(b)].
The most unstable mode is denoted by $(m_*, k_{y*})$ and satisfies $\sigma_{m_*} (k_{y*}) = \max_{k_y} \sigma_{m_*} (k_y)$.

To study the instability that emerges as $\beta_\mathrm{amp}$ increases, we numerically diagonalize $\mathbf{S}(k_y)$ at system size $(L_x, L_y) = (40, 80)$ and mean density $\bar{\rho} = 1/2$.
We first calculate the largest eigenvalues $\sigma_{m_*}(k_y)$ for several values of $\beta_\mathrm{amp}$ and fixed $\beta_\mathrm{mean} = 0.75$.
Note that $\beta_\mathrm{mean} < 1 / T_c$, with $T_c$ the critical temperature for a homogeneous profile.
Results are displayed on Fig.~\ref{fig_linear}(b).
They show that instability of the uniform density state appears for $\beta_\mathrm{amp} > \beta_p \simeq 0.08$, where $\beta_p$ is the phase boundary based on linear stability.
Further, the most unstable wavenumber $k_{y*}$ is nonzero near the stability limit, indicating that periodic density modulation will appear spontaneously along the $y$ axis, in sharp contrast with the homogeneous temperature case.

In Fig.~\ref{fig_linear}(c), we show the mode corresponding to the most unstable eigenvalue at $\beta_\mathrm{amp} = 0.08$.
More specifically, the eigenvector $\bm{v}$ that satisfies $\mathbf{S}(k_{y*}) \bm{v} = \sigma_{m_*} (k_{y*}) \bm{v}$ is transformed to a lattice-space representation as $\phi_{\bm{j}} \propto \mathrm{Re} [\sum_{k_x} v_{k_x} e^{i (k_x j_x + k_{y*} j_y)}]$ and plotted in Fig.~\ref{fig_linear}(c) with grayscale.
The result shows that this mode indeed features periodic density modulation along the $y$ axis, particularly in the low-temperature region with $T_{\bm{j}} < T_c$.
Furthermore, as shown with a streamline plot in Fig.~\ref{fig_linear}(c), we find that the linearized current obtained by Eq.~\eqref{eq_linearized_current} exhibits periodic circulations, reminiscent of convection cells in fluid dynamics~\cite{Chandrasekhar1961, Ahlers2009}.

The phase diagram in Fig.~\ref{fig_linear}(d) is mapped by calculating the critical amplitude $\beta_p$ as a function of the mean inverse temperature $\beta_{\mathrm{mean}}$.
This boundary separates the uniform density state from the periodic convective phase predicted by linear theory.
We find that $\beta_p + \beta_{\mathrm{mean}} \ge 1/T_c$, indicating that convection only emerges when the local temperature drops below $T_c$ in some region of the system.
Finite-size analysis for $L_x \in [40, 120]$ at a fixed aspect ratio $L_y/L_x = 2$ suggests that the phase boundary approaches the line $\beta_{\mathrm{mean}} + \beta_{\mathrm{amp}} = 1/T_c$ in the large-system limit (see Sec.~\ref{app_sizedep} of the SM).

\paragraph{Nonlinear simulations.}

To confirm the phase diagram predicted by the linear stability analysis, we numerically simulate the model described by Eqs.~\eqref{eq_model} and \eqref{eq_model_current} for the temperature profile~\eqref{eq_temperature_profile_example} [Fig.~\ref{fig_linear}(a)], $\bar{\rho} = 1 / 2$, and system size $(L_x,L_y) = (40,80)$.
We consider two kinds of initial states:
\begin{enumerate}
    \item[(I)]  a uniform density state supplemented with small noise: $\rho_{\bm{j}} (0) = \bar{\rho} + 0.001 \eta_{\bm{j}}$ with $\eta_{\bm{j}}$ taken independently from a standard normal distribution, neglecting small stochastic errors $(L_x L_y)^{-1} \sum_{\bm{j}} \rho_{\bm{j}} (0) - \bar{\rho} = O(10^{-5} \text{--} 10^{-4})$,

    \item[(II)] a completely segregated state along the $y$ axis: $\rho_{\bm{j}} (0) = 1$ for $0 \leq j_y < L_y / 2$ and $\rho_{\bm{j}} (0) = 0$ for $L_y / 2 \leq j_y < L_y$.
\end{enumerate}
We use a time step $\Delta t = 0.01$ and apply the Euler method for time integration of Eq.~\eqref{eq_model}.
See Ref.~\cite{code} for sample simulation codes, which can be executed on Google Colab to generate simulation videos.

\begin{figure}[t]
    \centering
    \includegraphics[scale=1.2]{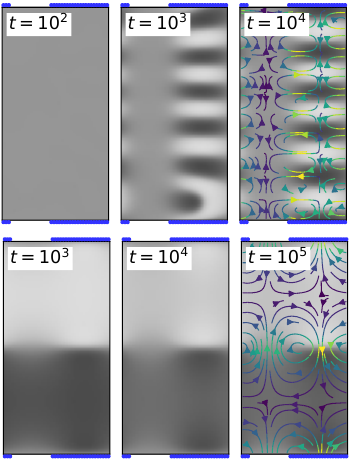}
    \caption{Dynamics of convective phase separation.
    Time evolution of the density field $\rho_{\bm{j}}(t)$ (grayscale) starting from a uniform initial state (I) (top panel) and a segregated state (II) (bottom panel).
    Blue dots indicate $T_{\bm{j}} < T_c$.
    Simulations are performed at $\beta_{\mathrm{mean}} = 0.83$ and $\beta_{\mathrm{amp}} = 0.09$, where the uniform state is linearly unstable. Colored arrows represent the steady-state current pattern overlaid on the final density field.
    Parameters: $(L_x, L_y) = (40, 80)$ and $\bar{\rho} = 1/2$.}
    \label{fig_mf_dynamics}
\end{figure}

Figure~\ref{fig_mf_dynamics} illustrates the evolution of the density field $\rho_{\bm{j}}(t)$ from the initial configurations (I) and (II), at $\beta_{\mathrm{mean}} = 0.83$ and $\beta_{\mathrm{amp}}= 0.09$.
These parameters lie within the linearly unstable regime [see Fig.~\ref{fig_linear}(d)].
See Sec.~\ref{app_dynamics} of the SM for additional parameters.
Consistent with our linear stability analysis, periodic phase separation emerges regardless of initial conditions.
The resulting density modulations are mainly located within the subcritical region $T_{\bm{j}} < T_c$ (blue dots, Fig.~\ref{fig_mf_dynamics}).

Starting from a uniform state (I), the system evolves into a steady state characterized by multiple convection cells.
Finite-size analysis for $L_x \in [40, 120]$ reveals that the number of cells grows sublinearly with system size, though a clear asymptotic scaling remains elusive (see Sec.~\ref{app_sizedep} of the SM).
In contrast, the initially segregated state (II) relaxes into a single high-density cluster without fragmentation, demonstrating that the final pattern selection is sensitive to the initial condition.
Crucially, for both initial conditions, the steady-state current exhibits a circulating, convection-cell-like pattern that qualitatively resembles the most unstable mode predicted by the linear theory [Fig.~\ref{fig_linear}(c)].
For the uniform initial state, the selected wavelength agrees with the prediction quantitatively near the onset of instability, while it shifts toward smaller wavenumbers away from the onset (see Sec.~\ref{app_stripenum} of the SM).
This circulating current pattern forms a robust and distinctive feature of the convecting phase.

We have checked that the detailed shape of the temperature profile is not essential: a triangle-wave inverse-temperature profile with the same mean and amplitude yields qualitatively identical convective phase separation.
We also performed simulations under radial temperature profiles to probe the generality of the behavior: we observed a spoke-like pattern or a single macroscopic domain depending on the profile, in both of which circulating currents appear (see Sec.~\ref{app_radial} of the SM).

The observed convective phase separation can be understood by considering the spatial variation of the currents.
In the low-temperature region ($j_x \sim 3 L_x / 4$), attractive interactions represented by the final term in Eq.~\eqref{eq_model_current} drive currents that segregate the density along the $y$ axis.
In the higher-temperature region ($j_x \sim L_x / 2$), normal diffusion, represented by the first two terms in Eq.~\eqref{eq_model_current}, becomes effective, generating currents that oppose the density gradient along the $x$ axis.
These two effects in distinct spatial regions are balanced globally to produce the circulating current pattern.

\begin{figure}[t]
    \centering
    \includegraphics[scale=1]{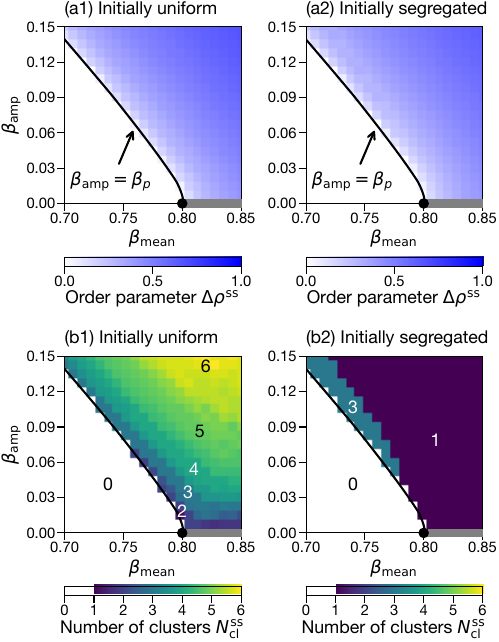}
    \caption{Steady-state phase diagram.
    (a) Heatmap of the order parameter $\Delta \rho^\mathrm{ss}$ for (I) uniform and (II) segregated initial states.
    The black line denotes the phase boundary predicted by linear stability analysis [cf. Fig.~\ref{fig_linear}(d)].
    (b) Heatmap of the number of high-density clusters $N_{\mathrm{cl}}^\mathrm{ss}$ for each initial state;
    digits indicate the cluster count within each region.
    Results for the uniform initial state [(a1) and (b1)] are averaged over 10 independent noise realizations.
    Parameters: $(L_x, L_y) = (40, 80)$ and $\bar{\rho} = 1/2$.}
    \label{fig_mf_steady}
\end{figure}

We next examine the dependence of the steady-state density field $\rho_{\bm{j}}^\mathrm{ss}$ on $\beta_\mathrm{mean}$ and $\beta_\mathrm{amp}$.
Practically, we consider that the steady state is reached when $t = 10^6$ or $\max_{\bm{j}} \{ |\partial_t \rho_{\bm{j}} (t)| \} < 10^{-10}$. Figures~\ref{fig_mf_steady}(a1) and \ref{fig_mf_steady}(a2) present the phase diagrams for the uniform (I) and segregated (II) initial states, respectively.
The heatmap displays the order parameter $\Delta \rho^\mathrm{ss} := \max_{\bm{j}} \rho_{\bm{j}}^\mathrm{ss} - \min_{\bm{j}} \rho_{\bm{j}}^\mathrm{ss}$, quantifying the amplitude of density modulation.
The onset of $\Delta \rho^\mathrm{ss} > 0$ shows excellent agreement with the predicted linearly unstable region, suggesting that the most unstable linear mode [Fig.~\ref{fig_linear}(c)] governs the steady-state transition; we stress that this pattern selection by the most unstable mode holds near the onset, whereas nonlinear effects drive the steady-state modulation toward smaller wavenumbers farther from it (see Sec.~\ref{app_stripenum} of the SM).

Figures~\ref{fig_mf_steady}(b1) and \ref{fig_mf_steady}(b2) show the steady-state number of high-density clusters, $N_\mathrm{cl}^\mathrm{ss}$.
We define a cluster as a connected component of at least 10 sites satisfying $\rho_{\bm{j}}^\mathrm{ss} > \bar{\rho} + 10^{-3}$.
We used the Python package scipy.ndimage.label~\cite{Virtanen2020} with the default structure parameter (i.e., the nearest neighbors regarded as connected) for detection of the connected components.
These results confirm that the number of convection cells depends on the initial condition; specifically, a single cluster ($N_\mathrm{cl}^\mathrm{ss} = 1$) persists for the segregated initial state (II) as one moves away from the transition boundary.

\begin{figure}[t]
    \centering
    \includegraphics[scale=1]{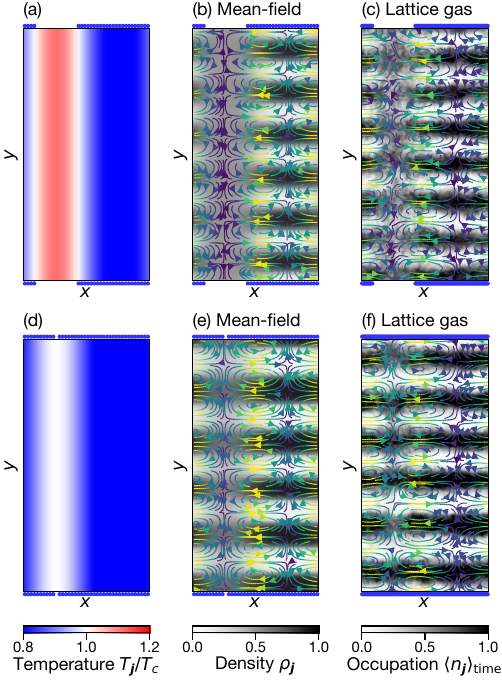}
    \caption{
    Comparison between the mean-field dynamics [Eqs.~\eqref{eq_model} and \eqref{eq_model_current}] and the stochastic lattice gas in Ref.~\cite{MS1}.
    (a) Temperature profile $T_{\bm{j}}$ for $\beta_{\mathrm{mean}} = 1.1/T_c$ and $\beta_{\mathrm{amp}} = 0.2/T_c$, used for panels (b) and (c).
    (b) Steady-state density field and currents for the mean-field model.
    (c) Time-averaged occupation field and currents for the stochastic model \cite{MS1}.
    (d--f) Corresponding results for $\beta_{\mathrm{mean}} = 1.2/T_c$ and $\beta_{\mathrm{amp}} = 0.2/T_c$.
    Blue dots indicate $T_{\bm{j}} < T_c$.
    Parameters:  $(L_x, L_y) = (40, 80)$ for mean-field,
    $(L_x, L_y) = (164, 328)$ for lattice gas,
    $\bar{\rho} = 1/2$ in both cases.}
    \label{fig_mf_comparison}
\end{figure}

Finally, we compare the steady states of the deterministic mean-field model [Eqs.~\eqref{eq_model} and \eqref{eq_model_current}] with the stochastic lattice gas from which it is derived \cite{MS1}.
We consider $\bar{\rho} = 1/2$ and a fixed aspect ratio $L_y/L_x = 2$, using the temperature profile in Eq.~\eqref{eq_temperature_profile_example} [Fig.~\ref{fig_mf_comparison}(a)].
Note that $T_c = 1.25$ in the mean-field model and $T_c \approx 0.567$ for the lattice gas \cite{Onsager1944}.

Figures~\ref{fig_mf_comparison}(b) and \ref{fig_mf_comparison}(c) display the steady-state density and currents for $\beta_{\mathrm{mean}} = 1.1/T_c$ and $\beta_{\mathrm{amp}} = 0.2/T_c$.
To clarify the current patterns in the stochastic model, we plot the time-averaged occupation field $\langle n_{\bm{j}} \rangle_{\mathrm{time}}$ and a spatially coarse-grained current (see Sec.~\ref{app_MC} of the SM).
Similar results for $\beta_{\mathrm{mean}}=1.2/T_c$ and $\beta_{\mathrm{amp}} = 0.2/T_c$ are shown in Figs.~\ref{fig_mf_comparison}(e) and \ref{fig_mf_comparison}(f).
To stabilize the density modulations against fluctuations, a larger system size $(L_x, L_y) = (164, 328)$ is used in the stochastic case.

Despite differences in microscopic detail, both models exhibit qualitatively similar periodic density modulations and convective currents.
This agreement demonstrates that the mean-field linear stability analysis captures the essential physics of the stochastic lattice gas.
Consequently, the deterministic framework provides a robust basis for further exploring the phase behavior of attractive particles in non-isothermal environments.
However, the two models differ in their sensitivity to initial conditions: the deterministic dynamics can be trapped in states dependent on the initial conditions, whereas fluctuations in the lattice gas remove this dependence (see Sec.~\ref{app_initdep} of the SM).

\paragraph{Discussion and Outlook.}
We have shown that convective phase separation in an inhomogeneous temperature field is governed by a linear instability that leads to robust, circulating steady-state currents.
Our central contribution is a deterministic mean-field description of the convective phase separation realized in a class of attractive lattice gases under temperature gradients, which pinpoints the linear instability and the circulating currents behind the patterned steady states.
In contrast to Rayleigh--B\'enard convection, which requires gravity, the convection cells here emerge from diffusive transport with attractive interactions.

An interesting direction is the role of the interaction sign.
While the attractive coupling considered here drives phase separation, a nearest-neighbor repulsion in the present single-component model would instead favor checkerboard-like sublattice ordering near half filling, as in the antiferromagnetic Ising model.
The situation is different for repulsive binary mixtures such as the Widom--Rowlinson lattice gas, in which only unlike species repel~\cite{Widom1970}, leading to macroscopic phase separation at high densities~\cite{Dickman1995}.
Driven versions of this model and related coarse-grained models exhibit rich pattern formation, including stripe reorientation~\cite{Dickman2018, Lavrentovich2021, Freire2026}.
Exploring whether the present framework can capture such phenomena would be a step toward a more unifying picture.
Looking forward, studying multi-component extensions in which thermal and chemical gradients compete is another natural direction.

\paragraph{Acknowledgments.}
K.A. thanks K.~Kawaguchi and N.~Nakagawa for insightful comments.
F.H. thanks R.K.P.~Zia for useful discussions.
K.A. acknowledges support by the RIKEN Information systems division for the use of the Supercomputer HOKUSAI BigWaterfall2.
This work was supported by JSPS KAKENHI Grant Numbers JP26K00052 and JP26K17111 (to K.A.).
Claude Opus and Fable were used to improve coding and readability.

\bibliography{ref_meanfield}

\begin{thebibliography}{47}%
\makeatletter
\providecommand \@ifxundefined [1]{%
 \@ifx{#1\undefined}
}%
\providecommand \@ifnum [1]{%
 \ifnum #1\expandafter \@firstoftwo
 \else \expandafter \@secondoftwo
 \fi
}%
\providecommand \@ifx [1]{%
 \ifx #1\expandafter \@firstoftwo
 \else \expandafter \@secondoftwo
 \fi
}%
\providecommand \natexlab [1]{#1}%
\providecommand \enquote  [1]{``#1''}%
\providecommand \bibnamefont  [1]{#1}%
\providecommand \bibfnamefont [1]{#1}%
\providecommand \citenamefont [1]{#1}%
\providecommand \href@noop [0]{\@secondoftwo}%
\providecommand \href [0]{\begingroup \@sanitize@url \@href}%
\providecommand \@href[1]{\@@startlink{#1}\@@href}%
\providecommand \@@href[1]{\endgroup#1\@@endlink}%
\providecommand \@sanitize@url [0]{\catcode `\\12\catcode `\$12\catcode
  `\&12\catcode `\#12\catcode `\^12\catcode `\_12\catcode `\%12\relax}%
\providecommand \@@startlink[1]{}%
\providecommand \@@endlink[0]{}%
\providecommand \url  [0]{\begingroup\@sanitize@url \@url }%
\providecommand \@url [1]{\endgroup\@href {#1}{\urlprefix }}%
\providecommand \urlprefix  [0]{URL }%
\providecommand \Eprint [0]{\href }%
\providecommand \doibase [0]{https://doi.org/}%
\providecommand \selectlanguage [0]{\@gobble}%
\providecommand \bibinfo  [0]{\@secondoftwo}%
\providecommand \bibfield  [0]{\@secondoftwo}%
\providecommand \translation [1]{[#1]}%
\providecommand \BibitemOpen [0]{}%
\providecommand \bibitemStop [0]{}%
\providecommand \bibitemNoStop [0]{.\EOS\space}%
\providecommand \EOS [0]{\spacefactor3000\relax}%
\providecommand \BibitemShut  [1]{\csname bibitem#1\endcsname}%
\let\auto@bib@innerbib\@empty
\bibitem [{\citenamefont {Cross}\ and\ \citenamefont
  {Hohenberg}(1993)}]{Cross1993}%
  \BibitemOpen
  \bibfield  {author} {\bibinfo {author} {\bibfnamefont {M.~C.}\ \bibnamefont
  {Cross}}\ and\ \bibinfo {author} {\bibfnamefont {P.~C.}\ \bibnamefont
  {Hohenberg}},\ }\bibfield  {title} {\bibinfo {title} {{Pattern formation
  outside of equilibrium}},\ }\href {https://doi.org/10.1103/RevModPhys.65.851}
  {\bibfield  {journal} {\bibinfo  {journal} {Rev. Mod. Phys.}\ }\textbf
  {\bibinfo {volume} {65}},\ \bibinfo {pages} {851} (\bibinfo {year}
  {1993})}\BibitemShut {NoStop}%
\bibitem [{\citenamefont {Cahn}\ and\ \citenamefont
  {Hilliard}(1958)}]{Cahn1958}%
  \BibitemOpen
  \bibfield  {author} {\bibinfo {author} {\bibfnamefont {J.~W.}\ \bibnamefont
  {Cahn}}\ and\ \bibinfo {author} {\bibfnamefont {J.~E.}\ \bibnamefont
  {Hilliard}},\ }\bibfield  {title} {\bibinfo {title} {{Free energy of a
  nonuniform system. I. interfacial free energy}},\ }\href
  {https://doi.org/10.1063/1.1744102} {\bibfield  {journal} {\bibinfo
  {journal} {J. Chem. Phys.}\ }\textbf {\bibinfo {volume} {28}},\ \bibinfo
  {pages} {258} (\bibinfo {year} {1958})}\BibitemShut {NoStop}%
\bibitem [{\citenamefont {Kawasaki}(1966)}]{Kawasaki1966}%
  \BibitemOpen
  \bibfield  {author} {\bibinfo {author} {\bibfnamefont {K.}~\bibnamefont
  {Kawasaki}},\ }\bibfield  {title} {\bibinfo {title} {{Diffusion Constants
  near the Critical Point for Time-Dependent Ising Models. I}},\ }\href
  {https://doi.org/10.1103/PhysRev.145.224} {\bibfield  {journal} {\bibinfo
  {journal} {Phys. Rev.}\ }\textbf {\bibinfo {volume} {145}},\ \bibinfo {pages}
  {224} (\bibinfo {year} {1966})}\BibitemShut {NoStop}%
\bibitem [{\citenamefont {Koch}\ \emph {et~al.}(1983)\citenamefont {Koch},
  \citenamefont {Desai},\ and\ \citenamefont {Abraham}}]{Koch1983}%
  \BibitemOpen
  \bibfield  {author} {\bibinfo {author} {\bibfnamefont {S.~W.}\ \bibnamefont
  {Koch}}, \bibinfo {author} {\bibfnamefont {R.~C.}\ \bibnamefont {Desai}},\
  and\ \bibinfo {author} {\bibfnamefont {F.~F.}\ \bibnamefont {Abraham}},\
  }\bibfield  {title} {\bibinfo {title} {{Dynamics of phase separation in
  two-dimensional fluids: Spinodal decomposition}},\ }\href
  {https://doi.org/10.1103/physreva.27.2152} {\bibfield  {journal} {\bibinfo
  {journal} {Phys. Rev. A}\ }\textbf {\bibinfo {volume} {27}},\ \bibinfo
  {pages} {2152} (\bibinfo {year} {1983})}\BibitemShut {NoStop}%
\bibitem [{\citenamefont {Bray}(1994)}]{Bray1994}%
  \BibitemOpen
  \bibfield  {author} {\bibinfo {author} {\bibfnamefont {A.~J.}\ \bibnamefont
  {Bray}},\ }\bibfield  {title} {\bibinfo {title} {{Theory of phase-ordering
  kinetics}},\ }\href {https://doi.org/10.1080/00018739400101505} {\bibfield
  {journal} {\bibinfo  {journal} {Adv. Phys.}\ }\textbf {\bibinfo {volume}
  {43}},\ \bibinfo {pages} {357} (\bibinfo {year} {1994})}\BibitemShut
  {NoStop}%
\bibitem [{\citenamefont {Onuki}(1997)}]{Onuki1997}%
  \BibitemOpen
  \bibfield  {author} {\bibinfo {author} {\bibfnamefont {A.}~\bibnamefont
  {Onuki}},\ }\bibfield  {title} {\bibinfo {title} {{Phase transitions of
  fluids in shear flow}},\ }\href {https://doi.org/10.1088/0953-8984/9/29/001}
  {\bibfield  {journal} {\bibinfo  {journal} {J. Phys. Condens. Matter}\
  }\textbf {\bibinfo {volume} {9}},\ \bibinfo {pages} {6119} (\bibinfo {year}
  {1997})}\BibitemShut {NoStop}%
\bibitem [{\citenamefont {Tanaka}(2000)}]{Tanaka2000}%
  \BibitemOpen
  \bibfield  {author} {\bibinfo {author} {\bibfnamefont {H.}~\bibnamefont
  {Tanaka}},\ }\bibfield  {title} {\bibinfo {title} {{Viscoelastic phase
  separation}},\ }\href {https://doi.org/10.1088/0953-8984/12/15/201}
  {\bibfield  {journal} {\bibinfo  {journal} {J. Phys. Condens. Matter}\
  }\textbf {\bibinfo {volume} {12}},\ \bibinfo {pages} {R207} (\bibinfo {year}
  {2000})}\BibitemShut {NoStop}%
\bibitem [{\citenamefont {Rubinstein}\ and\ \citenamefont
  {Colby}(2003)}]{Rubinstein2003}%
  \BibitemOpen
  \bibfield  {author} {\bibinfo {author} {\bibfnamefont {M.}~\bibnamefont
  {Rubinstein}}\ and\ \bibinfo {author} {\bibfnamefont {R.~H.}\ \bibnamefont
  {Colby}},\ }\href@noop {} {\emph {\bibinfo {title} {{Polymer Physics}}}}\
  (\bibinfo  {publisher} {Oxford University Press},\ \bibinfo {address}
  {Oxford},\ \bibinfo {year} {2003})\BibitemShut {NoStop}%
\bibitem [{\citenamefont {Cabral}\ and\ \citenamefont
  {Higgins}(2018)}]{Cabral2018}%
  \BibitemOpen
  \bibfield  {author} {\bibinfo {author} {\bibfnamefont {J.~T.}\ \bibnamefont
  {Cabral}}\ and\ \bibinfo {author} {\bibfnamefont {J.~S.}\ \bibnamefont
  {Higgins}},\ }\bibfield  {title} {\bibinfo {title} {{Spinodal nanostructures
  in polymer blends: On the validity of the Cahn-Hilliard length scale
  prediction}},\ }\href {https://doi.org/10.1016/j.progpolymsci.2018.03.003}
  {\bibfield  {journal} {\bibinfo  {journal} {Prog. Polym. Sci.}\ }\textbf
  {\bibinfo {volume} {81}},\ \bibinfo {pages} {1} (\bibinfo {year}
  {2018})}\BibitemShut {NoStop}%
\bibitem [{\citenamefont {Hohenberg}\ and\ \citenamefont
  {Halperin}(1977)}]{Hohenberg1977}%
  \BibitemOpen
  \bibfield  {author} {\bibinfo {author} {\bibfnamefont {P.~C.}\ \bibnamefont
  {Hohenberg}}\ and\ \bibinfo {author} {\bibfnamefont {B.~I.}\ \bibnamefont
  {Halperin}},\ }\bibfield  {title} {\bibinfo {title} {{Theory of dynamic
  critical phenomena}},\ }\href {https://doi.org/10.1103/RevModPhys.49.435}
  {\bibfield  {journal} {\bibinfo  {journal} {Rev. Mod. Phys.}\ }\textbf
  {\bibinfo {volume} {49}},\ \bibinfo {pages} {435} (\bibinfo {year}
  {1977})}\BibitemShut {NoStop}%
\bibitem [{\citenamefont {Vicsek}\ and\ \citenamefont
  {Zafeiris}(2012)}]{Vicsek2012}%
  \BibitemOpen
  \bibfield  {author} {\bibinfo {author} {\bibfnamefont {T.}~\bibnamefont
  {Vicsek}}\ and\ \bibinfo {author} {\bibfnamefont {A.}~\bibnamefont
  {Zafeiris}},\ }\bibfield  {title} {\bibinfo {title} {{Collective motion}},\
  }\href {https://doi.org/10.1016/j.physrep.2012.03.004} {\bibfield  {journal}
  {\bibinfo  {journal} {Phys. Rep.}\ }\textbf {\bibinfo {volume} {517}},\
  \bibinfo {pages} {71} (\bibinfo {year} {2012})}\BibitemShut {NoStop}%
\bibitem [{\citenamefont {Chaté}(2020)}]{Chate2020}%
  \BibitemOpen
  \bibfield  {author} {\bibinfo {author} {\bibfnamefont {H.}~\bibnamefont
  {Chaté}},\ }\bibfield  {title} {\bibinfo {title} {{Dry Aligning Dilute
  Active Matter}},\ }\href
  {https://doi.org/10.1146/annurev-conmatphys-031119-050752} {\bibfield
  {journal} {\bibinfo  {journal} {Annu. Rev. Condens. Matter Phys.}\ }\textbf
  {\bibinfo {volume} {11}},\ \bibinfo {pages} {189} (\bibinfo {year}
  {2020})}\BibitemShut {NoStop}%
\bibitem [{\citenamefont {Fily}\ and\ \citenamefont
  {Marchetti}(2012)}]{Fily2012}%
  \BibitemOpen
  \bibfield  {author} {\bibinfo {author} {\bibfnamefont {Y.}~\bibnamefont
  {Fily}}\ and\ \bibinfo {author} {\bibfnamefont {M.~C.}\ \bibnamefont
  {Marchetti}},\ }\bibfield  {title} {\bibinfo {title} {{Athermal phase
  separation of self-propelled particles with no alignment}},\ }\href
  {https://doi.org/10.1103/PhysRevLett.108.235702} {\bibfield  {journal}
  {\bibinfo  {journal} {Phys. Rev. Lett.}\ }\textbf {\bibinfo {volume} {108}},\
  \bibinfo {pages} {235702} (\bibinfo {year} {2012})}\BibitemShut {NoStop}%
\bibitem [{\citenamefont {Cates}\ and\ \citenamefont
  {Tailleur}(2015)}]{Cates2015}%
  \BibitemOpen
  \bibfield  {author} {\bibinfo {author} {\bibfnamefont {M.~E.}\ \bibnamefont
  {Cates}}\ and\ \bibinfo {author} {\bibfnamefont {J.}~\bibnamefont
  {Tailleur}},\ }\bibfield  {title} {\bibinfo {title} {{Motility-Induced Phase
  Separation}},\ }\href
  {https://doi.org/10.1146/annurev-conmatphys-031214-014710} {\bibfield
  {journal} {\bibinfo  {journal} {Annu. Rev. Condens. Matter Phys.}\ }\textbf
  {\bibinfo {volume} {6}},\ \bibinfo {pages} {219} (\bibinfo {year}
  {2015})}\BibitemShut {NoStop}%
\bibitem [{\citenamefont {Tjhung}\ \emph {et~al.}(2018)\citenamefont {Tjhung},
  \citenamefont {Nardini},\ and\ \citenamefont {Cates}}]{Tjhung2018}%
  \BibitemOpen
  \bibfield  {author} {\bibinfo {author} {\bibfnamefont {E.}~\bibnamefont
  {Tjhung}}, \bibinfo {author} {\bibfnamefont {C.}~\bibnamefont {Nardini}},\
  and\ \bibinfo {author} {\bibfnamefont {M.~E.}\ \bibnamefont {Cates}},\
  }\bibfield  {title} {\bibinfo {title} {{Cluster Phases and Bubbly Phase
  Separation in Active Fluids: Reversal of the Ostwald Process}},\ }\href
  {https://doi.org/10.1103/PhysRevX.8.031080} {\bibfield  {journal} {\bibinfo
  {journal} {Phys. Rev. X}\ }\textbf {\bibinfo {volume} {8}},\ \bibinfo {pages}
  {031080} (\bibinfo {year} {2018})}\BibitemShut {NoStop}%
\bibitem [{\citenamefont {Shi}\ \emph {et~al.}(2020)\citenamefont {Shi},
  \citenamefont {Fausti}, \citenamefont {Chaté}, \citenamefont {Nardini},\
  and\ \citenamefont {Solon}}]{Shi2020}%
  \BibitemOpen
  \bibfield  {author} {\bibinfo {author} {\bibfnamefont {X.-Q.}\ \bibnamefont
  {Shi}}, \bibinfo {author} {\bibfnamefont {G.}~\bibnamefont {Fausti}},
  \bibinfo {author} {\bibfnamefont {H.}~\bibnamefont {Chaté}}, \bibinfo
  {author} {\bibfnamefont {C.}~\bibnamefont {Nardini}},\ and\ \bibinfo {author}
  {\bibfnamefont {A.}~\bibnamefont {Solon}},\ }\bibfield  {title} {\bibinfo
  {title} {{Self-Organized Critical Coexistence Phase in Repulsive Active
  Particles}},\ }\href {https://doi.org/10.1103/PhysRevLett.125.168001}
  {\bibfield  {journal} {\bibinfo  {journal} {Phys. Rev. Lett.}\ }\textbf
  {\bibinfo {volume} {125}},\ \bibinfo {pages} {168001} (\bibinfo {year}
  {2020})}\BibitemShut {NoStop}%
\bibitem [{\citenamefont {Cates}\ and\ \citenamefont
  {Nardini}(2025)}]{Cates2025}%
  \BibitemOpen
  \bibfield  {author} {\bibinfo {author} {\bibfnamefont {M.~E.}\ \bibnamefont
  {Cates}}\ and\ \bibinfo {author} {\bibfnamefont {C.}~\bibnamefont
  {Nardini}},\ }\bibfield  {title} {\bibinfo {title} {{Active phase separation:
  new phenomenology from non-equilibrium physics}},\ }\href
  {https://doi.org/10.1088/1361-6633/add278} {\bibfield  {journal} {\bibinfo
  {journal} {Rep. Prog. Phys.}\ }\textbf {\bibinfo {volume} {88}},\ \bibinfo
  {pages} {056601} (\bibinfo {year} {2025})}\BibitemShut {NoStop}%
\bibitem [{\citenamefont {Keller}\ and\ \citenamefont
  {Segel}(1970)}]{Keller1970}%
  \BibitemOpen
  \bibfield  {author} {\bibinfo {author} {\bibfnamefont {E.~F.}\ \bibnamefont
  {Keller}}\ and\ \bibinfo {author} {\bibfnamefont {L.~A.}\ \bibnamefont
  {Segel}},\ }\bibfield  {title} {\bibinfo {title} {{Initiation of slime mold
  aggregation viewed as an instability}},\ }\href
  {https://doi.org/10.1016/0022-5193(70)90092-5} {\bibfield  {journal}
  {\bibinfo  {journal} {J. Theor. Biol.}\ }\textbf {\bibinfo {volume} {26}},\
  \bibinfo {pages} {399} (\bibinfo {year} {1970})}\BibitemShut {NoStop}%
\bibitem [{\citenamefont {Weyer}\ \emph {et~al.}(2025)\citenamefont {Weyer},
  \citenamefont {Muramatsu},\ and\ \citenamefont {Frey}}]{Weyer2025}%
  \BibitemOpen
  \bibfield  {author} {\bibinfo {author} {\bibfnamefont {H.}~\bibnamefont
  {Weyer}}, \bibinfo {author} {\bibfnamefont {D.}~\bibnamefont {Muramatsu}},\
  and\ \bibinfo {author} {\bibfnamefont {E.}~\bibnamefont {Frey}},\ }\bibfield
  {title} {\bibinfo {title} {{Chemotaxis-induced phase separation}},\ }\href
  {https://doi.org/10.1103/2933-45qc} {\bibfield  {journal} {\bibinfo
  {journal} {Phys. Rev. Lett.}\ }\textbf {\bibinfo {volume} {135}},\ \bibinfo
  {pages} {208402} (\bibinfo {year} {2025})}\BibitemShut {NoStop}%
\bibitem [{\citenamefont {Cotton}\ \emph {et~al.}(2022)\citenamefont {Cotton},
  \citenamefont {Golestanian},\ and\ \citenamefont
  {Agudo-Canalejo}}]{Cotton2022}%
  \BibitemOpen
  \bibfield  {author} {\bibinfo {author} {\bibfnamefont {M.~W.}\ \bibnamefont
  {Cotton}}, \bibinfo {author} {\bibfnamefont {R.}~\bibnamefont
  {Golestanian}},\ and\ \bibinfo {author} {\bibfnamefont {J.}~\bibnamefont
  {Agudo-Canalejo}},\ }\bibfield  {title} {\bibinfo {title} {{Catalysis-Induced
  Phase Separation and Autoregulation of Enzymatic Activity}},\ }\href
  {https://doi.org/10.1103/PhysRevLett.129.158101} {\bibfield  {journal}
  {\bibinfo  {journal} {Phys. Rev. Lett.}\ }\textbf {\bibinfo {volume} {129}},\
  \bibinfo {pages} {158101} (\bibinfo {year} {2022})}\BibitemShut {NoStop}%
\bibitem [{\citenamefont {Corberi}\ \emph {et~al.}(1998)\citenamefont
  {Corberi}, \citenamefont {Gonnella},\ and\ \citenamefont
  {Lamura}}]{Corberi1998}%
  \BibitemOpen
  \bibfield  {author} {\bibinfo {author} {\bibfnamefont {F.}~\bibnamefont
  {Corberi}}, \bibinfo {author} {\bibfnamefont {G.}~\bibnamefont {Gonnella}},\
  and\ \bibinfo {author} {\bibfnamefont {A.}~\bibnamefont {Lamura}},\
  }\bibfield  {title} {\bibinfo {title} {{Spinodal Decomposition of Binary
  Mixtures in Uniform Shear Flow}},\ }\href
  {https://doi.org/10.1103/PhysRevLett.81.3852} {\bibfield  {journal} {\bibinfo
   {journal} {Phys. Rev. Lett.}\ }\textbf {\bibinfo {volume} {81}},\ \bibinfo
  {pages} {3852} (\bibinfo {year} {1998})}\BibitemShut {NoStop}%
\bibitem [{\citenamefont {Davis}\ \emph {et~al.}(2025)\citenamefont {Davis},
  \citenamefont {A.},\ and\ \citenamefont {Sen~Gupta}}]{bbhy-gmsp}%
  \BibitemOpen
  \bibfield  {author} {\bibinfo {author} {\bibfnamefont {D.}~\bibnamefont
  {Davis}}, \bibinfo {author} {\bibfnamefont {P.}~\bibnamefont {A.}},\ and\
  \bibinfo {author} {\bibfnamefont {B.}~\bibnamefont {Sen~Gupta}},\ }\bibfield
  {title} {\bibinfo {title} {Phase separation and rheology of segregating
  binary fluid under shear},\ }\href {https://doi.org/10.1103/bbhy-gmsp}
  {\bibfield  {journal} {\bibinfo  {journal} {Phys. Rev. E}\ }\textbf {\bibinfo
  {volume} {111}},\ \bibinfo {pages} {065414} (\bibinfo {year}
  {2025})}\BibitemShut {NoStop}%
\bibitem [{\citenamefont {Katz}\ \emph {et~al.}(1984)\citenamefont {Katz},
  \citenamefont {Lebowitz},\ and\ \citenamefont {Spohn}}]{Katz1984}%
  \BibitemOpen
  \bibfield  {author} {\bibinfo {author} {\bibfnamefont {S.}~\bibnamefont
  {Katz}}, \bibinfo {author} {\bibfnamefont {J.~L.}\ \bibnamefont {Lebowitz}},\
  and\ \bibinfo {author} {\bibfnamefont {H.}~\bibnamefont {Spohn}},\ }\bibfield
   {title} {\bibinfo {title} {{Nonequilibrium steady states of stochastic
  lattice gas models of fast ionic conductors}},\ }\href
  {https://doi.org/10.1007/BF01018556} {\bibfield  {journal} {\bibinfo
  {journal} {J. Stat. Phys.}\ }\textbf {\bibinfo {volume} {34}},\ \bibinfo
  {pages} {497} (\bibinfo {year} {1984})}\BibitemShut {NoStop}%
\bibitem [{\citenamefont {Schmittmann}\ and\ \citenamefont
  {Zia}(1995)}]{Schmittmann1995}%
  \BibitemOpen
  \bibfield  {author} {\bibinfo {author} {\bibfnamefont {B.}~\bibnamefont
  {Schmittmann}}\ and\ \bibinfo {author} {\bibfnamefont {R.~K.~P.}\
  \bibnamefont {Zia}},\ }\bibfield  {title} {\bibinfo {title} {Statistical
  mechanics of driven diffusive systems},\ }in\ \href@noop {} {\emph {\bibinfo
  {booktitle} {Phase Transitions and Critical Phenomena}}},\ Vol.~\bibinfo
  {volume} {17}\ (\bibinfo  {publisher} {Academic Press},\ \bibinfo {address}
  {San Diego, CA},\ \bibinfo {year} {1995})\BibitemShut {NoStop}%
\bibitem [{\citenamefont {Marro}\ and\ \citenamefont
  {Dickman}(1999)}]{Marro1999}%
  \BibitemOpen
  \bibfield  {author} {\bibinfo {author} {\bibfnamefont {J.}~\bibnamefont
  {Marro}}\ and\ \bibinfo {author} {\bibfnamefont {R.}~\bibnamefont
  {Dickman}},\ }\href@noop {} {\emph {\bibinfo {title} {{Nonequilibrium Phase
  Transitions in Lattice Models}}}}\ (\bibinfo  {publisher} {Cambridge
  University Press},\ \bibinfo {address} {Cambridge, UK},\ \bibinfo {year}
  {1999})\BibitemShut {NoStop}%
\bibitem [{\citenamefont {Pleimling}\ \emph {et~al.}(2010)\citenamefont
  {Pleimling}, \citenamefont {Schmittmann},\ and\ \citenamefont
  {Zia}}]{Pleimling2010}%
  \BibitemOpen
  \bibfield  {author} {\bibinfo {author} {\bibfnamefont {M.}~\bibnamefont
  {Pleimling}}, \bibinfo {author} {\bibfnamefont {B.}~\bibnamefont
  {Schmittmann}},\ and\ \bibinfo {author} {\bibfnamefont {R.~K.~P.}\
  \bibnamefont {Zia}},\ }\bibfield  {title} {\bibinfo {title} {{Convection
  cells induced by spontaneous symmetry breaking}},\ }\href
  {https://doi.org/10.1209/0295-5075/89/50001} {\bibfield  {journal} {\bibinfo
  {journal} {EPL}\ }\textbf {\bibinfo {volume} {89}},\ \bibinfo {pages} {50001}
  (\bibinfo {year} {2010})}\BibitemShut {NoStop}%
\bibitem [{\citenamefont {Li}\ and\ \citenamefont {Pleimling}(2012)}]{Li2012}%
  \BibitemOpen
  \bibfield  {author} {\bibinfo {author} {\bibfnamefont {L.}~\bibnamefont
  {Li}}\ and\ \bibinfo {author} {\bibfnamefont {M.}~\bibnamefont {Pleimling}},\
  }\bibfield  {title} {\bibinfo {title} {{Formation of nonequilibrium modulated
  phases under local energy input}},\ }\href
  {https://doi.org/10.1209/0295-5075/98/30004} {\bibfield  {journal} {\bibinfo
  {journal} {EPL}\ }\textbf {\bibinfo {volume} {98}},\ \bibinfo {pages} {30004}
  (\bibinfo {year} {2012})}\BibitemShut {NoStop}%
\bibitem [{\citenamefont {Van~den Brande}\ \emph {et~al.}(2026)\citenamefont
  {Van~den Brande}, \citenamefont {Adachi},\ and\ \citenamefont
  {Huveneers}}]{MS1}%
  \BibitemOpen
  \bibfield  {author} {\bibinfo {author} {\bibfnamefont {M.}~\bibnamefont
  {Van~den Brande}}, \bibinfo {author} {\bibfnamefont {K.}~\bibnamefont
  {Adachi}},\ and\ \bibinfo {author} {\bibfnamefont {F.}~\bibnamefont
  {Huveneers}},\ }\bibfield  {title} {\bibinfo {title} {{Convection patterns in
  nonequilibrium Kawasaki dynamics at low temperature}},\ }\href
  {https://doi.org/10.1103/r6d8-bcd1} {\bibfield  {journal} {\bibinfo
  {journal} {Phys. Rev. E}\ }\textbf {\bibinfo {volume} {114}},\ \bibinfo
  {pages} {014147} (\bibinfo {year} {2026})}\BibitemShut {NoStop}%
\bibitem [{\citenamefont {Ball}\ and\ \citenamefont {Essery}(1990)}]{Ball1990}%
  \BibitemOpen
  \bibfield  {author} {\bibinfo {author} {\bibfnamefont {R.~C.}\ \bibnamefont
  {Ball}}\ and\ \bibinfo {author} {\bibfnamefont {R.~L.~H.}\ \bibnamefont
  {Essery}},\ }\bibfield  {title} {\bibinfo {title} {{Spinodal decomposition
  and pattern formation near surfaces}},\ }\href
  {https://doi.org/10.1088/0953-8984/2/51/006} {\bibfield  {journal} {\bibinfo
  {journal} {J. Phys. Condens. Matter}\ }\textbf {\bibinfo {volume} {2}},\
  \bibinfo {pages} {10303} (\bibinfo {year} {1990})}\BibitemShut {NoStop}%
\bibitem [{\citenamefont {Furukawa}(1992)}]{Furukawa1992}%
  \BibitemOpen
  \bibfield  {author} {\bibinfo {author} {\bibfnamefont {H.}~\bibnamefont
  {Furukawa}},\ }\bibfield  {title} {\bibinfo {title} {{Phase separation by
  directional quenching and morphological transition}},\ }\href
  {https://doi.org/10.1016/0378-4371(92)90111-3} {\bibfield  {journal}
  {\bibinfo  {journal} {Physica A}\ }\textbf {\bibinfo {volume} {180}},\
  \bibinfo {pages} {128} (\bibinfo {year} {1992})}\BibitemShut {NoStop}%
\bibitem [{\citenamefont {Alt}\ and\ \citenamefont {Pawlow}(1992)}]{Alt1992}%
  \BibitemOpen
  \bibfield  {author} {\bibinfo {author} {\bibfnamefont {H.~W.}\ \bibnamefont
  {Alt}}\ and\ \bibinfo {author} {\bibfnamefont {I.}~\bibnamefont {Pawlow}},\
  }\bibfield  {title} {\bibinfo {title} {{A mathematical model of dynamics of
  non-isothermal phase separation}},\ }\href
  {https://doi.org/10.1016/0167-2789(92)90078-2} {\bibfield  {journal}
  {\bibinfo  {journal} {Physica D}\ }\textbf {\bibinfo {volume} {59}},\
  \bibinfo {pages} {389} (\bibinfo {year} {1992})}\BibitemShut {NoStop}%
\bibitem [{\citenamefont {Araki}\ and\ \citenamefont
  {Tanaka}(2004)}]{Araki2004}%
  \BibitemOpen
  \bibfield  {author} {\bibinfo {author} {\bibfnamefont {T.}~\bibnamefont
  {Araki}}\ and\ \bibinfo {author} {\bibfnamefont {H.}~\bibnamefont {Tanaka}},\
  }\bibfield  {title} {\bibinfo {title} {{Hydrodynamic delocalization of phase
  separation in a locally cooled fluid mixture}},\ }\href
  {https://doi.org/10.1209/epl/i2003-10073-7} {\bibfield  {journal} {\bibinfo
  {journal} {EPL}\ }\textbf {\bibinfo {volume} {65}},\ \bibinfo {pages} {214}
  (\bibinfo {year} {2004})}\BibitemShut {NoStop}%
\bibitem [{\citenamefont {Jaiswal}\ \emph {et~al.}(2013)\citenamefont
  {Jaiswal}, \citenamefont {Puri},\ and\ \citenamefont {Binder}}]{Jaiswal2013}%
  \BibitemOpen
  \bibfield  {author} {\bibinfo {author} {\bibfnamefont {P.~K.}\ \bibnamefont
  {Jaiswal}}, \bibinfo {author} {\bibfnamefont {S.}~\bibnamefont {Puri}},\ and\
  \bibinfo {author} {\bibfnamefont {K.}~\bibnamefont {Binder}},\ }\bibfield
  {title} {\bibinfo {title} {{Phase separation in thin films: Effect of
  temperature gradients}},\ }\href
  {https://doi.org/10.1209/0295-5075/103/66003} {\bibfield  {journal} {\bibinfo
   {journal} {EPL}\ }\textbf {\bibinfo {volume} {103}},\ \bibinfo {pages}
  {66003} (\bibinfo {year} {2013})}\BibitemShut {NoStop}%
\bibitem [{\citenamefont {Chandrasekhar}(1961)}]{Chandrasekhar1961}%
  \BibitemOpen
  \bibfield  {author} {\bibinfo {author} {\bibfnamefont {S.}~\bibnamefont
  {Chandrasekhar}},\ }\href@noop {} {\emph {\bibinfo {title} {Hydrodynamic and
  Hydromagnetic Stability}}}\ (\bibinfo  {publisher} {Clarendon Press},\
  \bibinfo {address} {Oxford},\ \bibinfo {year} {1961})\BibitemShut {NoStop}%
\bibitem [{\citenamefont {Ahlers}\ \emph {et~al.}(2009)\citenamefont {Ahlers},
  \citenamefont {Grossmann},\ and\ \citenamefont {Lohse}}]{Ahlers2009}%
  \BibitemOpen
  \bibfield  {author} {\bibinfo {author} {\bibfnamefont {G.}~\bibnamefont
  {Ahlers}}, \bibinfo {author} {\bibfnamefont {S.}~\bibnamefont {Grossmann}},\
  and\ \bibinfo {author} {\bibfnamefont {D.}~\bibnamefont {Lohse}},\ }\bibfield
   {title} {\bibinfo {title} {{Heat transfer and large scale dynamics in
  turbulent Rayleigh-Bénard convection}},\ }\href
  {https://doi.org/10.1103/RevModPhys.81.503} {\bibfield  {journal} {\bibinfo
  {journal} {Rev. Mod. Phys.}\ }\textbf {\bibinfo {volume} {81}},\ \bibinfo
  {pages} {503} (\bibinfo {year} {2009})}\BibitemShut {NoStop}%
\bibitem [{\citenamefont {Turing}(1952)}]{Turing1952}%
  \BibitemOpen
  \bibfield  {author} {\bibinfo {author} {\bibfnamefont {A.~M.}\ \bibnamefont
  {Turing}},\ }\bibfield  {title} {\bibinfo {title} {{The chemical basis of
  morphogenesis}},\ }\href {https://doi.org/10.1098/rstb.1952.0012} {\bibfield
  {journal} {\bibinfo  {journal} {Philos. Trans. R. Soc.}\ }\textbf {\bibinfo
  {volume} {237}},\ \bibinfo {pages} {37} (\bibinfo {year} {1952})}\BibitemShut
  {NoStop}%
\bibitem [{\citenamefont {Murray}(2002)}]{Murray2002}%
  \BibitemOpen
  \bibfield  {author} {\bibinfo {author} {\bibfnamefont {J.~D.}\ \bibnamefont
  {Murray}},\ }\href@noop {} {\emph {\bibinfo {title} {{Mathematical biology
  II: Spatial models and biomedical applications}}}}\ (\bibinfo  {publisher}
  {Springer},\ \bibinfo {address} {New York},\ \bibinfo {year}
  {2002})\BibitemShut {NoStop}%
\bibitem [{\citenamefont {Penrose}(1991)}]{Penrose1991}%
  \BibitemOpen
  \bibfield  {author} {\bibinfo {author} {\bibfnamefont {O.}~\bibnamefont
  {Penrose}},\ }\bibfield  {title} {\bibinfo {title} {{A mean-field equation of
  motion for the dynamic Ising model}},\ }\href
  {https://doi.org/10.1007/bf01029993} {\bibfield  {journal} {\bibinfo
  {journal} {J. Stat. Phys.}\ }\textbf {\bibinfo {volume} {63}},\ \bibinfo
  {pages} {975} (\bibinfo {year} {1991})}\BibitemShut {NoStop}%
\bibitem [{\citenamefont {Oono}\ and\ \citenamefont {Puri}(1988)}]{Oono1988}%
  \BibitemOpen
  \bibfield  {author} {\bibinfo {author} {\bibfnamefont {Y.}~\bibnamefont
  {Oono}}\ and\ \bibinfo {author} {\bibfnamefont {S.}~\bibnamefont {Puri}},\
  }\bibfield  {title} {\bibinfo {title} {{Study of phase-separation dynamics by
  use of cell dynamical systems. I. Modeling}},\ }\href
  {https://doi.org/10.1103/physreva.38.434} {\bibfield  {journal} {\bibinfo
  {journal} {Phys. Rev. A}\ }\textbf {\bibinfo {volume} {38}},\ \bibinfo
  {pages} {434} (\bibinfo {year} {1988})}\BibitemShut {NoStop}%
\bibitem [{cod()}]{code}%
  \BibitemOpen
  \href {https://github.com/adachi24/convective-phase-separation} {\bibinfo
  {title}
  {https://github.com/adachi24/convective-phase-separation}}\BibitemShut
  {NoStop}%
\bibitem [{\citenamefont {Virtanen}\ \emph {et~al.}(2020)\citenamefont
  {Virtanen}, \citenamefont {Gommers}, \citenamefont {Oliphant}, \citenamefont
  {Haberland}, \citenamefont {Reddy}, \citenamefont {Cournapeau}, \citenamefont
  {Burovski}, \citenamefont {Peterson}, \citenamefont {Weckesser},
  \citenamefont {Bright}, \citenamefont {van~der Walt}, \citenamefont {Brett},
  \citenamefont {Wilson}, \citenamefont {Millman}, \citenamefont {Mayorov},
  \citenamefont {Nelson}, \citenamefont {Jones}, \citenamefont {Kern},
  \citenamefont {Larson}, \citenamefont {Carey}, \citenamefont {Polat},
  \citenamefont {Feng}, \citenamefont {Moore}, \citenamefont {VanderPlas},
  \citenamefont {Laxalde}, \citenamefont {Perktold}, \citenamefont {Cimrman},
  \citenamefont {Henriksen}, \citenamefont {Quintero}, \citenamefont {Harris},
  \citenamefont {Archibald}, \citenamefont {Ribeiro}, \citenamefont
  {Pedregosa}, \citenamefont {van Mulbregt},\ and\ \citenamefont {{SciPy 1.0
  Contributors}}}]{Virtanen2020}%
  \BibitemOpen
  \bibfield  {author} {\bibinfo {author} {\bibfnamefont {P.}~\bibnamefont
  {Virtanen}}, \bibinfo {author} {\bibfnamefont {R.}~\bibnamefont {Gommers}},
  \bibinfo {author} {\bibfnamefont {T.~E.}\ \bibnamefont {Oliphant}}, \bibinfo
  {author} {\bibfnamefont {M.}~\bibnamefont {Haberland}}, \bibinfo {author}
  {\bibfnamefont {T.}~\bibnamefont {Reddy}}, \bibinfo {author} {\bibfnamefont
  {D.}~\bibnamefont {Cournapeau}}, \bibinfo {author} {\bibfnamefont
  {E.}~\bibnamefont {Burovski}}, \bibinfo {author} {\bibfnamefont
  {P.}~\bibnamefont {Peterson}}, \bibinfo {author} {\bibfnamefont
  {W.}~\bibnamefont {Weckesser}}, \bibinfo {author} {\bibfnamefont
  {J.}~\bibnamefont {Bright}}, \bibinfo {author} {\bibfnamefont {S.~J.}\
  \bibnamefont {van~der Walt}}, \bibinfo {author} {\bibfnamefont
  {M.}~\bibnamefont {Brett}}, \bibinfo {author} {\bibfnamefont
  {J.}~\bibnamefont {Wilson}}, \bibinfo {author} {\bibfnamefont {K.~J.}\
  \bibnamefont {Millman}}, \bibinfo {author} {\bibfnamefont {N.}~\bibnamefont
  {Mayorov}}, \bibinfo {author} {\bibfnamefont {A.~R.~J.}\ \bibnamefont
  {Nelson}}, \bibinfo {author} {\bibfnamefont {E.}~\bibnamefont {Jones}},
  \bibinfo {author} {\bibfnamefont {R.}~\bibnamefont {Kern}}, \bibinfo {author}
  {\bibfnamefont {E.}~\bibnamefont {Larson}}, \bibinfo {author} {\bibfnamefont
  {C.~J.}\ \bibnamefont {Carey}}, \bibinfo {author} {\bibfnamefont
  {{\.I}.}~\bibnamefont {Polat}}, \bibinfo {author} {\bibfnamefont
  {Y.}~\bibnamefont {Feng}}, \bibinfo {author} {\bibfnamefont {E.~W.}\
  \bibnamefont {Moore}}, \bibinfo {author} {\bibfnamefont {J.}~\bibnamefont
  {VanderPlas}}, \bibinfo {author} {\bibfnamefont {D.}~\bibnamefont {Laxalde}},
  \bibinfo {author} {\bibfnamefont {J.}~\bibnamefont {Perktold}}, \bibinfo
  {author} {\bibfnamefont {R.}~\bibnamefont {Cimrman}}, \bibinfo {author}
  {\bibfnamefont {I.}~\bibnamefont {Henriksen}}, \bibinfo {author}
  {\bibfnamefont {E.~A.}\ \bibnamefont {Quintero}}, \bibinfo {author}
  {\bibfnamefont {C.~R.}\ \bibnamefont {Harris}}, \bibinfo {author}
  {\bibfnamefont {A.~M.}\ \bibnamefont {Archibald}}, \bibinfo {author}
  {\bibfnamefont {A.~H.}\ \bibnamefont {Ribeiro}}, \bibinfo {author}
  {\bibfnamefont {F.}~\bibnamefont {Pedregosa}}, \bibinfo {author}
  {\bibfnamefont {P.}~\bibnamefont {van Mulbregt}},\ and\ \bibinfo {author}
  {\bibnamefont {{SciPy 1.0 Contributors}}},\ }\bibfield  {title} {\bibinfo
  {title} {{SciPy 1.0: fundamental algorithms for scientific computing in
  Python}},\ }\href {https://doi.org/10.1038/s41592-019-0686-2} {\bibfield
  {journal} {\bibinfo  {journal} {Nat. Methods}\ }\textbf {\bibinfo {volume}
  {17}},\ \bibinfo {pages} {261} (\bibinfo {year} {2020})}\BibitemShut
  {NoStop}%
\bibitem [{\citenamefont {Onsager}(1944)}]{Onsager1944}%
  \BibitemOpen
  \bibfield  {author} {\bibinfo {author} {\bibfnamefont {L.}~\bibnamefont
  {Onsager}},\ }\bibfield  {title} {\bibinfo {title} {{Crystal Statistics. I. A
  Two-Dimensional Model with an Order-Disorder Transition}},\ }\href
  {https://doi.org/10.1103/PhysRev.65.117} {\bibfield  {journal} {\bibinfo
  {journal} {Phys. Rev.}\ }\textbf {\bibinfo {volume} {65}},\ \bibinfo {pages}
  {117} (\bibinfo {year} {1944})}\BibitemShut {NoStop}%
\bibitem [{\citenamefont {Widom}\ and\ \citenamefont
  {Rowlinson}(1970)}]{Widom1970}%
  \BibitemOpen
  \bibfield  {author} {\bibinfo {author} {\bibfnamefont {B.}~\bibnamefont
  {Widom}}\ and\ \bibinfo {author} {\bibfnamefont {J.~S.}\ \bibnamefont
  {Rowlinson}},\ }\bibfield  {title} {\bibinfo {title} {{New model for the
  study of liquid--vapor phase transitions}},\ }\href
  {https://doi.org/10.1063/1.1673203} {\bibfield  {journal} {\bibinfo
  {journal} {J. Chem. Phys.}\ }\textbf {\bibinfo {volume} {52}},\ \bibinfo
  {pages} {1670} (\bibinfo {year} {1970})}\BibitemShut {NoStop}%
\bibitem [{\citenamefont {Dickman}\ and\ \citenamefont
  {Stell}(1995)}]{Dickman1995}%
  \BibitemOpen
  \bibfield  {author} {\bibinfo {author} {\bibfnamefont {R.}~\bibnamefont
  {Dickman}}\ and\ \bibinfo {author} {\bibfnamefont {G.}~\bibnamefont
  {Stell}},\ }\bibfield  {title} {\bibinfo {title} {{Critical behavior of the
  Widom--Rowlinson lattice model}},\ }\href {https://doi.org/10.1063/1.468970}
  {\bibfield  {journal} {\bibinfo  {journal} {J. Chem. Phys.}\ }\textbf
  {\bibinfo {volume} {102}},\ \bibinfo {pages} {8674} (\bibinfo {year}
  {1995})}\BibitemShut {NoStop}%
\bibitem [{\citenamefont {Dickman}\ and\ \citenamefont
  {Zia}(2018)}]{Dickman2018}%
  \BibitemOpen
  \bibfield  {author} {\bibinfo {author} {\bibfnamefont {R.}~\bibnamefont
  {Dickman}}\ and\ \bibinfo {author} {\bibfnamefont {R.~K.~P.}\ \bibnamefont
  {Zia}},\ }\bibfield  {title} {\bibinfo {title} {Driven {W}idom-{R}owlinson
  lattice gas},\ }\href {https://doi.org/10.1103/PhysRevE.97.062126} {\bibfield
   {journal} {\bibinfo  {journal} {Phys. Rev. E}\ }\textbf {\bibinfo {volume}
  {97}},\ \bibinfo {pages} {062126} (\bibinfo {year} {2018})}\BibitemShut
  {NoStop}%
\bibitem [{\citenamefont {Lavrentovich}\ \emph {et~al.}(2021)\citenamefont
  {Lavrentovich}, \citenamefont {Dickman},\ and\ \citenamefont
  {Zia}}]{Lavrentovich2021}%
  \BibitemOpen
  \bibfield  {author} {\bibinfo {author} {\bibfnamefont {M.~O.}\ \bibnamefont
  {Lavrentovich}}, \bibinfo {author} {\bibfnamefont {R.}~\bibnamefont
  {Dickman}},\ and\ \bibinfo {author} {\bibfnamefont {R.~K.~P.}\ \bibnamefont
  {Zia}},\ }\bibfield  {title} {\bibinfo {title} {Microemulsions in the driven
  {W}idom-{R}owlinson lattice gas},\ }\href
  {https://doi.org/10.1103/PhysRevE.104.064135} {\bibfield  {journal} {\bibinfo
   {journal} {Phys. Rev. E}\ }\textbf {\bibinfo {volume} {104}},\ \bibinfo
  {pages} {064135} (\bibinfo {year} {2021})}\BibitemShut {NoStop}%
\bibitem [{\citenamefont {Freire~Oliveira}\ \emph {et~al.}(2026)\citenamefont
  {Freire~Oliveira}, \citenamefont {Dickman}, \citenamefont {Lavrentovich},\
  and\ \citenamefont {Zia}}]{Freire2026}%
  \BibitemOpen
  \bibfield  {author} {\bibinfo {author} {\bibfnamefont {G.~E.}\ \bibnamefont
  {Freire~Oliveira}}, \bibinfo {author} {\bibfnamefont {R.}~\bibnamefont
  {Dickman}}, \bibinfo {author} {\bibfnamefont {M.~O.}\ \bibnamefont
  {Lavrentovich}},\ and\ \bibinfo {author} {\bibfnamefont {R.~K.~P.}\
  \bibnamefont {Zia}},\ }\bibfield  {title} {\bibinfo {title} {Pattern
  formation in a coupled driven diffusive system},\ }\href
  {https://doi.org/10.1103/23pw-x6sd} {\bibfield  {journal} {\bibinfo
  {journal} {Phys. Rev. E}\ }\textbf {\bibinfo {volume} {113}},\ \bibinfo
  {pages} {014105} (\bibinfo {year} {2026})}\BibitemShut {NoStop}%
\end{thebibliography}%

\clearpage
\onecolumngrid 
\appendix

\begin{center}
    \textbf{\large Supplemental Material:
    Mean-Field Convective Phase Separation under Thermal Gradients} \\
    \vspace{10pt}
    Meander Van den Brande, François Huveneers and Kyosuke Adachi
\end{center}

\setcounter{equation}{0}
\setcounter{figure}{0}
\setcounter{table}{0}
\setcounter{section}{0}
\setcounter{page}{1}

\counterwithout{equation}{section}

\renewcommand{\theequation}{S\arabic{equation}}
\renewcommand{\thefigure}{S\arabic{figure}}
\renewcommand{\thesection}{S\arabic{section}}

\makeatletter
\renewcommand{\section}[1]{%
  \par\refstepcounter{section}%
  \vspace{10pt} 
  \begin{center}
    \textbf{S\arabic{section}. #1} 
  \end{center}
  \par\vspace{4pt} 
}
\makeatother

\section{Mean-Field Approximation of the Stochastic Model in Ref.~\cite{MS1}}
\label{app_MFA}

We derive the deterministic model equations Eqs.~\eqref{eq_model} and \eqref{eq_model_current} from the stochastic model in Ref.~\cite{MS1}, through a mean-field approximation.
This stochastic model is a Kawasaki dynamics of $N$ particles on a rectangular lattice of size $(L_x, L_y)$ with on-site exclusion, nearest-neighbor attractive interaction, and inhomogeneous temperature.
Below, $J$ represents the attractive interaction strength and $\gamma$ the intrinsic rate constant of hopping in Ref.~\cite{MS1}.
As in the main text, we let $\bm{j} = j_x \bm{e}_x + j_y \bm{e}_y$ be the two-dimensional coordinate.
The hopping rate of a particle at site $\bm{j}$ to a neighboring empty site $\bm{j} + \bm{e}_a$ ($a \in \{ x, y\}$) and the counterpart for the reverse process are defined as
\begin{equation}\label{appeq_rates}
    w_{\bm{j} \to \bm{j} + \bm{e}_a}  = \frac{\gamma}{2} \bigg[ 1 + \tanh \bigg( \frac{2 J \tilde{\delta}_{\bm{j}}^a}{T_{\bm{j} + \bm{e}_a / 2}} \bigg) \bigg],
    \qquad
    w_{\bm{j} + \bm{e}_a \to \bm{j}}  = \frac{\gamma}{2} \bigg[ 1 + \tanh \bigg( \frac{-2 J \tilde{\delta}_{\bm{j}}^a}{T_{\bm{j} + \bm{e}_a / 2}} \bigg) \bigg],
\end{equation}
where $\tilde{\delta}_{\bm{j}}^a$ is the change in the total number of neighboring particles after the particle jump in the $a$ direction:
\begin{equation}
\begin{aligned}
   \tilde{\delta}^x_{\bm{j}} & = n_{\bm{j} + 2 \bm{e}_x} + n_{\bm{j} + \bm{e}_x + \bm{e}_y} + n_{\bm{j} + \bm{e}_x - \bm{e}_y} - n_{\bm{j} + \bm{e}_y} - n_{\bm{j} - \bm{e}_y} - n_{\bm{j} - \bm{e}_x} ,\\
   \tilde{\delta}^y_{\bm{j}} & = n_{\bm{j} + 2 \bm{e}_y} + n_{\bm{j} + \bm{e}_y + \bm{e}_x} + n_{\bm{j} + \bm{e}_y - \bm{e}_x} - n_{\bm{j} + \bm{e}_x} - n_{\bm{j} - \bm{e}_x} - n_{\bm{j} - \bm{e}_y}.
   \label{appeq_deltahat}
\end{aligned}
\end{equation}
Here, $n_{\bm{j}} \in \{ 0, 1 \}$ is the occupation of site $\bm{j}$, and the set $\{ n_{\bm{j}} \}_{\bm{j}}$ specifies the particle configuration.
Throughout this work, we fix the temperature and time units by setting $J = 1/4$ and $\gamma = 2$, for which Eq.~\eqref{appeq_rates} reduces to $w_{\bm{j} \to \bm{j} + \bm{e}_a} = 1 + \tanh [ \tilde{\delta}_{\bm{j}}^a / (2 T_{\bm{j} + \bm{e}_a / 2}) ]$.
In these units, the transition rates can exceed unity; only the transition probabilities $w \Delta t$ per time step $\Delta t$ enter the Monte Carlo simulations, and $\Delta t$ is chosen such that the total transition probability does not exceed unity (see Sec.~\ref{app_MC}).

Writing the ensemble average at time $t$ as $\braket{\cdots}$, we obtain the dynamical equation for the ensemble-averaged particle density, $\rho_{\bm{j}} (t) := \braket{n_{\bm{j}}} \in [0, 1]$:
\begin{equation}
   \partial_t \rho_{\bm{j}} = -(J^x_{\bm{j}} - J^x_{\bm{j} - \bm{e}_x} + J^y_{\bm{j}} - J^y_{\bm{j} - \bm{e}_y}),
   \label{appeq_eoc}
\end{equation}
where the ensemble-averaged particle number current  $J_{\bm{j}}^a$ ($a \in \{ x, y \}$) is given as
\begin{equation}
    J^a_{\bm{j}} = \braket{n_{\bm{j}} (1 - n_{\bm{j} + \bm{e}_a}) w_{\bm{j} \to \bm{j} + \bm{e}_a} - n_{\bm{j} + \bm{e}_a} (1 - n_{\bm{j}}) w_{\bm{j} + \bm{e}_a \to \bm{j}}}.
    \label{appeq_current_exact}
\end{equation}

By a mean-field approximation that neglects all the microscopic correlations, Eq.~\eqref{appeq_current_exact} reduces to
\begin{align}
    J^a_{\bm{j}} & \approx \rho_{\bm{j}} (1 - \rho_{\bm{j} + \bm{e}_a}) \bigg[ 1 + \tanh \bigg( \frac{\braket{\tilde{\delta}_{\bm{j}}^a}}{2 T_{\bm{j} + \bm{e}_a / 2}} \bigg) \bigg]
    - \rho_{\bm{j} + \bm{e}_a} (1 - \rho_{\bm{j}}) \bigg[ 1 + \tanh \bigg( \frac{-\braket{\tilde{\delta}_{\bm{j}}^a}}{2 T_{\bm{j} + \bm{e}_a / 2}} \bigg) \bigg] \nonumber \\
    & = \rho_{\bm{j}} - \rho_{\bm{j} + \bm{e}_a} + (\rho_{\bm{j}} + \rho_{\bm{j} + \bm{e}_a} - 2 \rho_{\bm{j}} \rho_{\bm{j} + \bm{e}_a}) \tanh \bigg( \frac{\braket{\tilde{\delta}_{\bm{j}}^a}}{2 T_{\bm{j} + \bm{e}_a / 2}} \bigg).
    \label{appeq_current_mfa}
\end{align}
From Eq.~\eqref{appeq_deltahat}, we see that
\begin{equation}
\begin{aligned}
   \braket{\tilde{\delta}^x_{\bm{j}}} & = \rho_{\bm{j} + 2 \bm{e}_x} + \rho_{\bm{j} + \bm{e}_x + \bm{e}_y} + \rho_{\bm{j} + \bm{e}_x - \bm{e}_y} - \rho_{\bm{j} + \bm{e}_y} - \rho_{\bm{j} - \bm{e}_y} - \rho_{\bm{j} - \bm{e}_x}, \\
   \braket{\tilde{\delta}^y_{\bm{j}}} & = \rho_{\bm{j} + 2 \bm{e}_y} + \rho_{\bm{j} + \bm{e}_y + \bm{e}_x} + \rho_{\bm{j} + \bm{e}_y - \bm{e}_x} - \rho_{\bm{j} + \bm{e}_x} - \rho_{\bm{j} - \bm{e}_x} - \rho_{\bm{j} - \bm{e}_y}.
   \label{appeq_deltahatexp}
\end{aligned}
\end{equation}
Thus, the obtained mean-field dynamical equation [Eqs.~\eqref{appeq_eoc}, \eqref{appeq_current_mfa}, and \eqref{appeq_deltahatexp}] is equivalent to the model in the main text [Eqs.~\eqref{eq_model}--\eqref{eq_model_delta}] by writing $\braket{\tilde{\delta}^a_{\bm{j}}}$ as $\delta^a_{\bm{j}}$.

\section{Linear Stability in the Homogeneous Temperature Case}
\label{app_EPS}

\begin{figure}[t]
    \centering
    \includegraphics[scale=1]{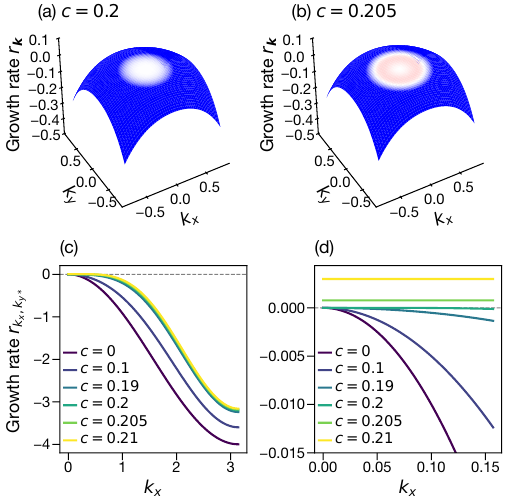}
    \caption{Growth rate of Fourier modes for the case of homogeneous temperature.
    (a), (b) Surface plot of the growth rate $r_{\bm{k}}$ as a function of $(k_x, k_y)$ at (a) $c = \bar{\rho} (1 - \bar{\rho}) / T = 0.2$ or (b) $c = 0.205$.
    Blue and red colors indicate negative and positive $r_{\bm{k}}$, respectively.
    (c) Growth rate $r_{k_x, k_{y*}}$, which is maximized for $k_y$, as a function of $k_x$.
    Different colors indicate different values of $c$.
    (d) Expanded view of panel (c).}
    \label{appfig_eps}
\end{figure}

We investigate numerically the linear stability of the homogeneous profile for a uniform temperature $T_{\bm{j}} = T$.
In this case, the linear evolution operator $\bf R$ in Eq.~\eqref{eq_linearized_dynamics} becomes diagonal in Fourier space:
\begin{equation}
     \partial_t \hat{\phi}_{\bm{k}} = r_{\bm{k}} \hat{\phi}_{\bm{k}},
     \label{eq_homogeneousT_linearized_dynamics}
\end{equation}
with a growth rate $r_{\bm k}$ derived from Eq.~\eqref{eq_Rkq}:
\begin{equation}\label{eq_homogeneousT_rk}
    r_{\bm{k}}  = -2 (2 - \cos k_x - \cos k_y)
     + 4 c \bigg[ \sin \frac{k_x}{2} \bigg( \sin \frac{3 k_x}{2} + 2 \sin \frac{k_x}{2} \cos k_y \bigg)
     + \sin \frac{k_y}{2} \bigg( \sin \frac{3 k_y}{2} + 2 \sin \frac{k_y}{2} \cos k_x \bigg) \bigg].
\end{equation}
Here, we have defined a coefficient parameter $c := \bar{\rho} (1 - \bar{\rho}) / T$.

The uniform density state is linearly unstable if $\max_{\bm{k}} r_{\bm{k}} > 0$, and the most unstable mode is specified by $\bm{k}_*$ that satisfies $r_{\bm{k}_*} = \max_{\bm{k}} r_{\bm{k}}$.
We numerically check the parameter dependence of $r_{\bm{k}}$.
We consider a sufficiently large system and regard $k_x$ and $k_y$ as continuous variables.
In Figs.~\ref{appfig_eps}(a) and \ref{appfig_eps}(b), we show the surface plots of $r_{\bm{k}}$ as a function of $(k_x, k_y)$ at $c = 0.2$ and $c = 0.205$, respectively.
The blue and red colors indicate negative and positive values of $r_{\bm{k}}$, respectively.
These figures show that unstable modes with positive $r_{\bm{k}}$ start to appear at $\bm{k} \approx \bm{0}$ as $c$ increases.

To see the $c$ dependence of the growth rate for the most unstable mode $\bm{k}_*$, we first maximize $r_{\bm{k}}$ with respect to $k_y$ and then visualize the $k_x$ dependence.
By writing the numerically optimized $k_y$ as $k_{y*}$ for each $k_x$, we plot $r_{k_x, k_{y*}}$ as a function of $k_x$ in Fig.~\ref{appfig_eps}(c), with the expanded view near $k_x = 0$ in Fig.~\ref{appfig_eps}(d).
We see that $r_{\bm{k}} < 0$ holds for $\bm{k} \neq \bm{0}$ as long as $0 < c < 0.2$.

\section{Explicit Expression for the Matrix $\mathbf{S}(k_y)$ in Eq.~\eqref{eq_S_matrix}}

We provide the explicit expression for the matrix $\mathbf{S}(k_y)$ in Eq.~\eqref{eq_S_matrix}, that is diagonalized to produce Fig.~\ref{fig_linear}.
The nonzero elements of the Fourier transformation $\hat{\beta}_{\bm{k}}$ are given by
\begin{equation}
    \hat{\beta}_{\bm{0}}  = L_x L_y \beta_\mathrm{mean},
    \qquad
    \hat{\beta}_{\pm q_0 \bm{e}_x}  = \pm i L_x L_y \beta_\mathrm{amp} / 2,
\end{equation}
where the smallest nonzero wavenumber along the $x$ axis is defined as $q_0 := 2 \pi / L_x$.
The nonzero elements of the $L_x \times L_x$ matrix $\mathbf{S}(k_y)$ are given by
\begin{equation}
\begin{aligned}
    S_{k_x k_x}(k_y) & = -2 (2 - \cos k_x - \cos k_y) \\
    & \quad + 4 \bar{\rho} (1 - \bar{\rho}) \beta_\mathrm{mean} \bigg[ \sin \frac{k_x}{2} \bigg( \sin \frac{3 k_x}{2} + 2 \sin \frac{k_x}{2} \cos k_y \bigg)
    + \sin \frac{k_y}{2} \bigg( \sin \frac{3 k_y}{2} + 2 \sin \frac{k_y}{2} \cos k_x \bigg) \bigg]
\end{aligned}
\end{equation}
and
\begin{equation}
\begin{aligned}
    S_{k_x, k_x \pm q_0}(k_y)  &= \mp 2 i \bar{\rho} (1 - \bar{\rho}) \beta_\mathrm{amp}  \\
    &\quad \times \bigg\{ \sin \frac{k_x}{2} \bigg[ \sin \frac{3 (k_x \pm q_0)}{2} + 2 \sin \frac{k_x \pm q_0}{2} \cos k_y \bigg]
    + \sin \frac{k_y}{2} \bigg[ \sin \frac{3 k_y}{2} + 2 \sin \frac{k_y}{2} \cos (k_x \pm q_0) \bigg] \bigg\}.
\end{aligned}
\end{equation}

\section{System Size Dependence}
\label{app_sizedep}

\begin{figure}[t]
    \centering
    \includegraphics[scale=1]{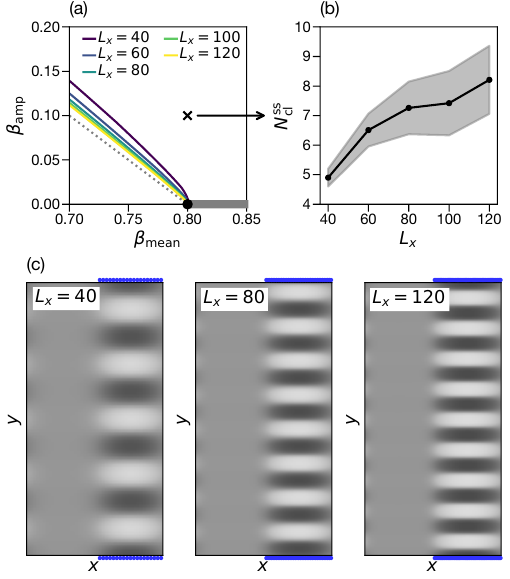}
    \caption{System size dependence of the phase boundary and steady pattern with a fixed aspect ratio $L_y / L_x = 2$ and $\bar\rho=1/2$.
    (a) Phase boundary based on linear stability for various $L_x$.
    The gray dotted line corresponds to $\min_{\bm{j}} T_{\bm{j}} = T_c$.
    (b) Steady-state cluster count $N_\mathrm{cl}^\mathrm{ss}$ at $\beta_\mathrm{mean} = 0.8$ and  $\beta_\mathrm{amp} = 0.1$ [cross in panel (a)].
    Dots and gray shadow indicate the mean and standard deviation over $100$ independent noise realizations in the initial state, respectively.
    (c) Typical steady patterns for $L_x = 40, 80, 120$.
    Blue dots indicate the low-temperature region where $T_{\bm{j}} < T_c$.}
    \label{fig_mf_sizedep}
\end{figure}

We examine the system-size dependence at fixed mean density $\bar{\rho} = 1/2$ and aspect ratio $L_y/L_x = 2$.
Figure~\ref{fig_mf_sizedep}(a) shows the linear stability boundary for $L_x \in [40, 120]$.
As the system size increases, the boundary approaches the line $\beta_{\mathrm{mean}} + \beta_{\mathrm{amp}} = 1/T_c = 0.8$ (gray dotted line).
This is consistent with the requirement that the local temperature must fall below $T_c$ at some spatial point to trigger convective phase separation.

Turning to the nonlinear dynamics, we investigate the scaling of the steady-state cluster count $N_{\mathrm{cl}}^{\mathrm{ss}}$ for the uniform initial state.
At $\beta_{\mathrm{mean}}=0.8$ and $\beta_{\mathrm{amp}} = 0.1$ [cross in Fig.~\ref{fig_mf_sizedep}(a)], we performed 100 independent simulations per system size using a time step $\Delta t = 0.05$ and the convergence criteria described in the main text.
Figure~\ref{fig_mf_sizedep}(b) shows the dependence of $N_{\mathrm{cl}}^{\mathrm{ss}}$ on $L_x$, with markers and shaded regions representing the mean and standard deviation, respectively.
While $N_{\mathrm{cl}}^{\mathrm{ss}}$ increases sublinearly with $L_x$, the asymptotic scaling remains unclear within the current simulation range.
We note that an analogous anomalous scaling of the number of patterns with system size has been reported in a related driven diffusive system~\cite{Freire2026}.
Typical steady-state density patterns for various system sizes are displayed in Fig.~\ref{fig_mf_sizedep}(c).

\section{Time Evolution: Further Parameters}
\label{app_dynamics}

\begin{figure}[t]
    \centering
    \includegraphics[scale=1]{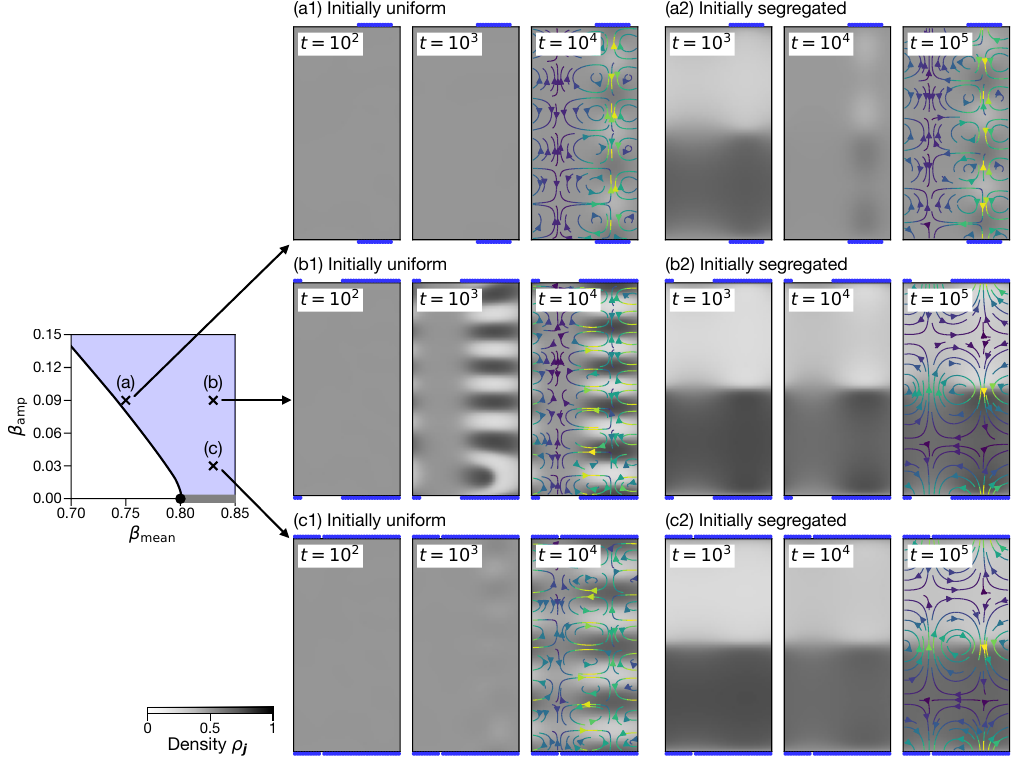}
    \caption{
    Dynamics of convective phase separation.
    Grayscale panels represent the time evolution of the density field $\rho_{\bm{j}} (t)$, starting from either the uniform state (I) or the segregated state (II).
    Parameters $(\beta_\mathrm{mean}, \beta_\mathrm{amp})$ are set to (a) $(0.75, 0.09)$, (b) $(0.83, 0.09)$, and (c) $(0.83, 0.03)$.
    Blue dots indicate low-temperature regions where $T_{\bm{j}} < T_c$.
    The time point is indicated at the top left of each panel; for the final time point, the density field is overlaid with the colored current pattern. As shown in the leftmost panel [reproduced from Fig.~\ref{fig_linear}(d)], these $(\beta_\mathrm{mean}, \beta_\mathrm{amp})$ values are selected from the region where the uniform state is linearly unstable.
    Other parameters are $(L_x, L_y) = (40, 80)$ and $\bar{\rho} = 1/2$.}
    \label{fig_mf_dynamics_app}
\end{figure}

To complement Fig.~\ref{fig_mf_dynamics} in the main text, we analyze the time evolution for other parameter values for $\beta_\mathrm{mean}$ and  $\beta_\mathrm{amp}$ within the convective region predicted by the linear stability analysis.
In Fig.~\ref{fig_mf_dynamics_app}, we show the time evolution of the density field $\rho_{\bm{j}} (t)$  starting from either the uniform state (I) or the segregated state (II) (see main text) at (a) $(\beta_\mathrm{mean}, \beta_\mathrm{amp}) = (0.75, 0.09)$, (b) $(0.83, 0.09)$, and (c) $(0.83, 0.03)$, where the uniform density state should be linearly unstable [see the left panel of Fig.~\ref{fig_mf_dynamics_app}, reproduced from Fig.~\ref{fig_linear}(d)].
As predicted by the linear stability analysis, we find that the periodic phase separation emerges when starting from the uniform density state with small noise [Figs.~\ref{fig_mf_dynamics_app}(a1), \ref{fig_mf_dynamics_app}(b1), and \ref{fig_mf_dynamics_app}(c1)].
We also see that phase separation appears largely in the low-temperature region with $T_{\bm{j}} < T_c$, indicated by blue dots in each panel of Fig.~\ref{fig_mf_dynamics_app}.
Each panel of the density field at the last time point is overlaid with the observed currents, exhibiting the periodic convective pattern essentially similar to that of the most unstable mode found in the linear stability analysis [Fig.~\ref{fig_linear}(c)].

The density modulation and current pattern also spontaneously appear when starting from the segregated state at $(\beta_\mathrm{mean}, \beta_\mathrm{amp}) = (0.75, 0.09)$ [Fig.~\ref{fig_mf_dynamics_app}(a2)], suggesting that periodic pattern formation is an intrinsic feature of the model~\eqref{eq_model}.
On the other hand, with the segregated initial state at $(\beta_\mathrm{mean}, \beta_\mathrm{amp}) = (0.83, 0.09)$ or $(0.83, 0.03)$, the single high-density cluster remains without dividing into multiple clusters after a long time [Figs.~\ref{fig_mf_dynamics_app}(b2) and \ref{fig_mf_dynamics_app}(c2)].
Comparing them to the counterparts with the uniform initial state [Figs.~\ref{fig_mf_dynamics_app}(b1) and \ref{fig_mf_dynamics_app}(c1)], we see that the steady pattern depends on the initial state.
However, even when only a single high-density cluster exists after a long time, the convective current pattern still appears as a distinctive feature of the model~\eqref{eq_model}, as indicated by the last configurations in Figs.~\ref{fig_mf_dynamics_app}(b2) and \ref{fig_mf_dynamics_app}(c2).

\section{Steady-State Number of Clusters Compared with the Linear-Stability Prediction}
\label{app_stripenum}

\begin{figure}[t]
    \centering
    \includegraphics[scale=1]{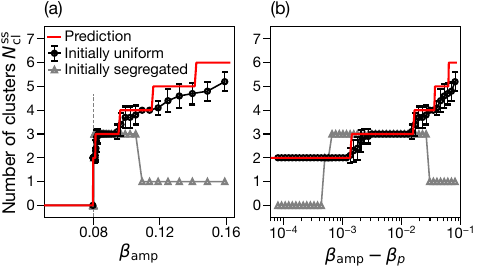}
    \caption{Steady-state number of clusters $N_\mathrm{cl}^\mathrm{ss}$ compared with the linear-stability prediction.
    $N_\mathrm{cl}^\mathrm{ss}$ is plotted as a function of $\beta_\mathrm{amp}$ at fixed $\beta_\mathrm{mean} = 0.75$, starting from the uniform initial state (I) (black circles: mean $\pm$ standard deviation over 10 noise realizations) and from the segregated initial state (II) (gray triangles), shown (a) on a linear scale and (b) as a function of the distance to the stability limit, $\beta_\mathrm{amp} - \beta_p$, on a logarithmic scale.
    The red line shows the number of periods of the most unstable mode along the $y$ axis, $N_\mathrm{cl}^\mathrm{pred} = k_{y*} L_y / (2 \pi)$, and the vertical dashed line in (a) marks the stability limit $\beta_p \simeq 0.08$.
    Parameters: $(L_x, L_y) = (40, 80)$ and $\bar{\rho} = 1/2$.}
    \label{appfig_stripenum}
\end{figure}

To quantify to what extent the most unstable linear mode governs the steady-state pattern away from the stability limit, we compare the steady-state number of clusters $N_\mathrm{cl}^\mathrm{ss}$ (see the main text for the definition) with the number of periods of the most unstable mode along the $y$ axis, $N_\mathrm{cl}^\mathrm{pred} := k_{y*} L_y / (2 \pi)$, at fixed $\beta_\mathrm{mean} = 0.75$ as a function of $\beta_\mathrm{amp}$ (Fig.~\ref{appfig_stripenum}).
The steady states are obtained with the same numerical protocol as in the main text, for both the uniform initial state (I) with 10 independent noise realizations and the segregated initial state (II).
Near the stability limit $\beta_p \simeq 0.08$, the cluster number for the uniform initial state follows the prediction, $N_\mathrm{cl}^\mathrm{ss} = N_\mathrm{cl}^\mathrm{pred} = 2$ and $3$, within the corresponding windows of $\beta_\mathrm{amp}$, except in the immediate vicinity of the predicted steps.
For $\beta_\mathrm{amp} \gtrsim 1.2 \beta_p$, however, $N_\mathrm{cl}^\mathrm{ss}$ starts to lag behind $N_\mathrm{cl}^\mathrm{pred}$: the observed steps are shifted toward larger $\beta_\mathrm{amp}$, and the deviation grows with the distance from the stability limit; additionally, different noise realizations select different cluster numbers depending on $\beta_\mathrm{amp}$, as indicated by the error bars in Fig.~\ref{appfig_stripenum}.
For the segregated initial state, the deviation is stronger, and the steady state locks into a single cluster for $\beta_\mathrm{amp} \gtrsim 0.11$.
Thus, the most unstable mode quantitatively determines the selected pattern only near the stability limit, while nonlinear effects drive the steady-state modulation toward smaller wavenumbers farther from it.
The qualitative structure, i.e., high-density clusters accompanied by circulating currents, persists whenever a patterned steady state forms.

\section{Radial Temperature Profiles}
\label{app_radial}

\begin{figure}[t]
    \centering
    \includegraphics[scale=1]{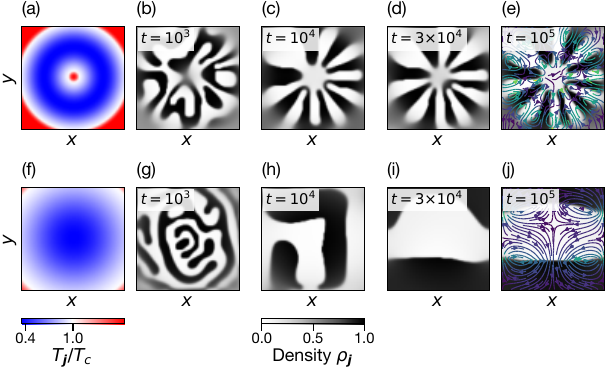}
    \caption{Mean-field dynamics under radial temperature profiles.
    (a) Temperature field $T_{\bm{j}}/T_c$ of the Mexican-hat profile, and [(b)--(e)] the time evolution of the density field $\rho_{\bm{j}}(t)$ (grayscale) at the indicated times, starting from the uniform initial state (I).
    [(f)--(j)] The same for the single-dip profile.
    The final panels ($t = 10^5$) are overlaid with the current pattern.
    Parameters: $(L_x, L_y) = (100, 100)$, $\bar{\rho} = 1/2$, and $(T_\mathrm{mean}, T_\mathrm{amp}) = (0.88, 0.44) \simeq (0.70 \, T_c, 0.35 \, T_c)$.}
    \label{appfig_radial}
\end{figure}

To explore temperature profiles beyond those modulated along a single direction, we consider radial profiles similar to those studied for the stochastic model in Ref.~\cite{MS1}: $T_{\bm{j}} = T_\mathrm{mean} + T_\mathrm{amp} \varphi_{1, 2}(r)$, with
\begin{equation}
    \varphi_1(r) =
    \begin{cases}
        \cos(2\pi \alpha_1 r) & (\alpha_1 r \leq 1) \\
        1 & (\alpha_1 r > 1)
    \end{cases},
    \qquad
    \varphi_2(r) =
    \begin{cases}
        -\cos(2\pi \alpha_2 r) & (\alpha_2 r \leq 1/2) \\
        1 & (\alpha_2 r > 1/2)
    \end{cases},
    \label{appeq_radial}
\end{equation}
where $r$ is the distance from the center of the system in units of half the linear system size, and $\alpha_1 \simeq 0.88$ and $\alpha_2 \simeq 0.32$ ensure that $\varphi_1$ and $\varphi_2$ have zero spatial average; $\varphi_1$ produces a Mexican-hat-shaped temperature landscape and $\varphi_2$ a single central dip [Figs.~\ref{appfig_radial}(a) and \ref{appfig_radial}(f)].
Starting from the uniform initial state (I) at $(T_\mathrm{mean}, T_\mathrm{amp}) \simeq (0.70 \, T_c, 0.35 \, T_c)$, the resulting dynamics differ between the two profiles: the Mexican-hat profile develops a spoke-like pattern organized around the central high-temperature region, similar to the pattern observed in the stochastic model (Fig.~1 of Ref.~\cite{MS1}), whereas the single-dip profile produces a single macroscopic domain, in contrast to the nonstationary pattern observed in the stochastic model (Fig.~3 of Ref.~\cite{MS1}).
Nevertheless, the currents exhibit circulating patterns in both cases [Figs.~\ref{appfig_radial}(e) and \ref{appfig_radial}(j)], as observed in the cases with temperature variations along a single direction.

\section{Monte Carlo Simulation of the Lattice Gas Model}
\label{app_MC}

We follow Ref.~\cite{MS1} for the numerical implementation of the stochastic lattice gas model presented in Sec.~\ref{app_MFA}.
We write the total particle number as $N$.
In a single Monte Carlo step, we update the particle configuration as follows:
\begin{enumerate}[label=(\arabic*)]
    \item We randomly choose a particle.
    \item The chosen particle (at site $\bm{j}$) hops to one of the four neighboring empty sites (site $\bm{j}'$) with probability $w_{\bm{j} \to \bm{j}'} \Delta t$, where $\Delta t$ is a time step and the hopping rate $w_{\bm{j} \to \bm{j}'}$ is given by Eq.~\eqref{appeq_rates}.
    \item We repeat procedures (1) and (2) $N$ times.
\end{enumerate}

We used $\Delta t = 1 / 8$, $(L_x, L_y) = (164, 328)$, and $N = L_x L_y / 2 = 26896$ in simulations for Fig.~\ref{fig_mf_comparison}.
Note that, although individual rates $w_{\bm{j} \to \bm{j}'}$ can exceed unity in our units (Sec.~\ref{app_MFA}), the total transition probability per update satisfies $\sum_{\bm{j}' (\neq \bm{j})} w_{\bm{j} \to \bm{j}'} \Delta t \leq 4 \gamma \Delta t = 1$ for $\Delta t = 1/8$.
For the time-averaged occupation fields $\braket{n_{\bm{j}}}_\mathrm{time}$ plotted in Figs.~\ref{fig_mf_comparison}(c) and \ref{fig_mf_comparison}(f), we used $11$ configurations at different time points with an interval of $5 \times 10^5$ MC steps after relaxation of $4.5 \times 10^7$ MC steps, starting with a random particle configuration.
To obtain the current patterns in Figs.~\ref{fig_mf_comparison}(c) and \ref{fig_mf_comparison}(f), we first performed a time average of the observed currents over the temporal domain used to compute $\braket{n_{\bm{j}}}_\mathrm{time}$; we then applied a moving average with a $9 \times 9$ window in the lattice space to reduce noise.

\section{Dependence on Initial Conditions: Mean-Field Model versus Stochastic Lattice Gas}
\label{app_initdep}

\begin{figure}[t]
    \centering
    \includegraphics[width=\textwidth]{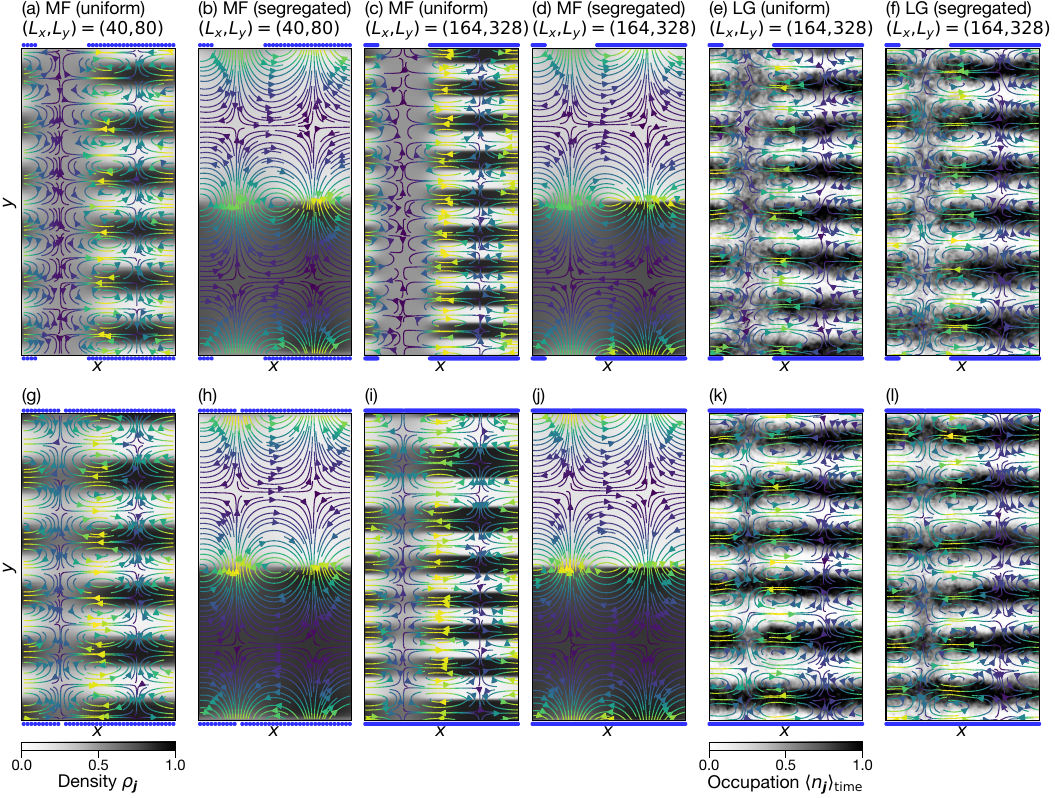}
    \caption{Dependence of the long-time behavior on the initial condition and system size, at the parameters for Fig.~\ref{fig_mf_comparison}: $(\beta_\mathrm{mean}, \beta_\mathrm{amp}) = (1.1/T_c, 0.2/T_c)$ [(a)--(f)] and $(1.2/T_c, 0.2/T_c)$ [(g)--(l)] with $\bar{\rho} = 1/2$.
    We show the density and currents of the mean-field model (MF) and the time-averaged occupation and coarse-grained currents of the stochastic lattice gas (LG), starting from the uniform or segregated initial condition at the system size indicated above each column.
    Data processing and rendering are the same as used in Fig.~\ref{fig_mf_comparison}.}
    \label{appfig_initdep}
\end{figure}

To examine the role of initial conditions in the two models, we repeat the comparison of Fig.~\ref{fig_mf_comparison} starting from the segregated initial state (II), for both the mean-field model and the stochastic lattice gas, together with mean-field simulations at the lattice-gas system size $(L_x, L_y) = (164, 328)$ (Fig.~\ref{appfig_initdep}).
Starting from the segregated state, the mean-field dynamics remains locked in a single-cluster state [Figs.~\ref{appfig_initdep}(b), \ref{appfig_initdep}(d), \ref{appfig_initdep}(h), and \ref{appfig_initdep}(j)], whereas the lattice gas escapes this configuration [Figs.~\ref{appfig_initdep}(f) and \ref{appfig_initdep}(l)] and reaches a periodic pattern similar to the one obtained with the uniform initial condition [Figs.~\ref{appfig_initdep}(e) and \ref{appfig_initdep}(k)]: fluctuations remove the initial-condition dependence characteristic of the deterministic dynamics.
In the lattice-gas dynamics, however, the number of clusters can fluctuate over long times and practically depends on the noise realization, as seen by comparing Figs.~\ref{appfig_initdep}(e) and \ref{appfig_initdep}(f).

Regarding the size of the convection cells, at the common system size $(L_x, L_y) = (164, 328)$ with the uniform initial condition, the mean-field model selects $9$ and $7$ clusters while the lattice gas selects $7$ and $6$ clusters at $\beta_\mathrm{mean} = 1.1/T_c$ and $\beta_\mathrm{mean} = 1.2/T_c$, respectively [Figs.~\ref{appfig_initdep}(c), \ref{appfig_initdep}(e), \ref{appfig_initdep}(i), and \ref{appfig_initdep}(k)].
Though this is a comparison with a single noise realization, the results suggest that the size of the convection cells depends on whether the dynamics is deterministic or stochastic even at the same system size.
We also note that the mean-field dynamics is slow especially at low temperatures, and Fig.~\ref{appfig_initdep}(i) can be a transient state that has not yet reached the steady state.

\end{document}